\documentclass[useAMS,usenatbib]{mn2e}
\usepackage{graphicx}
\usepackage{amsmath}
\usepackage{amssymb}
\usepackage{comment}
\usepackage{subfigure}

\include{defns}
\def\gsim{\;\rlap{\lower 2.5pt
 \hbox{$\sim$}}\raise 1.5pt\hbox{$>$}\;}
\def\lsim{\;\rlap{\lower 2.5pt
   \hbox{$\sim$}}\raise 1.5pt\hbox{$<$}\;}
\defcitealias{liu09a}{Liu et al. (2009a)}
\defcitealias{liu09b}{Liu et al (2009b)}

\newcommand{\tr}[1]{\textrm{#1}}
\newcommand{\ee}[1]{\times10^{#1}}
\newcommand{\dd}[2]{\frac{\textrm{d}#1}{\textrm{d}#2}}
\newcommand{\pp}[2]{\frac{\partial#1}{\partial#2}}

\title[Cosmic Ray Streaming in Clusters of Galaxies]{Cosmic Ray Streaming in Clusters of Galaxies}
\author[Wiener, Oh, \& Guo]{Joshua Wiener$^{1}$, S. Peng Oh$^{1}$, \& Fulai Guo$^{2}$ $^{3}$ \\
$^{1}$Department of Physics; University of California; Santa Barbara, CA 93106, USA.\\
$^{2}$Department of Astronomy \& Astrophysics, University of California, Santa Cruz, CA 95064, USA.\\
$^{3}$ETH Z\"urich, Institute for Astronomy, Wolfgang-Pauli-Strasse 27, CH-8093, Z\"urich, Switzerland}

\begin{document}
\bibliographystyle{mn2e}

\pagerange{000--000} \pubyear{0000}
\maketitle

\label{firstpage}

\begin{abstract}
The observed bimodality in radio luminosity in galaxy clusters is puzzling. We investigate the possibility that cosmic-ray (CR) streaming in the intra-cluster medium can `switch off' hadronically induced radio and gamma-ray emission. For self-confined CRs, this depends on the source of MHD wave damping: if only non-linear Landau damping operates, then CRs stream on the slow Alfv\'enic timescale, but if turbulent wave damping operates, super-Alfv\'enic streaming is possible. As turbulence increases, it promotes outward streaming more than it enables inward turbulent advection. Curiously, the CR flux is {\it independent} of $\nabla f$ (as long as it is non-zero) and depends only on plasma parameters; this enables radio halos with flat inferred CR profiles to turn off. We perform 1D time-dependent calculations of a radio mini-halo (Perseus) and giant radio halo (Coma) and find that both diminish in radio luminosity by an order of magnitude in several hundred Myr, given plausible estimates for the magnetic field in the outskirts of the cluster. Due to the energy dependence of CR streaming, spectral curvature develops, and radio halos turn off more slowly at low frequencies -- properties consistent with observations. Similarly, CR streaming rapidly turns off gamma-ray emission at the high-energies probed by Cherenkov telescopes, but not at the low energies probed by {\it Fermi}. CR mediated wave-heating of the ICM is unaffected, as it is dominated by $\sim$GeV CRs which stream Alfv\'enically.

%The scattering of cosmic ray protons traveling through a plasma is a topic of some interest.  The observed isotropy of cosmic ray synchrotron emission in our own galaxy suggests a long cosmic ray lifetime.  This confinement may be explained by scattering of the cosmic rays in the ISM, and so possible scattering mechanisms have been studied.  The primary scatterer is thought to be Alfv\'en waves generated by the cosmic rays themselves.  This self-confinement has been studied in detail (see \cite{farmer04}) and cosmic ray streaming speeds have been predicted.  We first summarize this analysis, and extend its results to cosmic rays streaming in galaxy clusters. We also ran a 1D cosmic ray simulation using a variation of the ZEUS3D code, adapted to track quantities as functions of CR momentum as well as determine the amount of radio synchrotron emission. From this we predict a time scale for the Perseus radio halo to switch off.
\end{abstract}


\section{Introduction}
\label{section:intro}

% para on general significance of CRs
% COMPARE WITH LONGER VERSION IN ATP10 PROPOSAL
Cosmic rays (CRs) in the intra-cluster medium (ICM) can arise from structure formation shocks \citep{miniati01,pfrommer08}, turbulent reacceleration of existing non-thermal particles \citep{brunetti07}, galactic winds and supernovae \citep{volk96}, and radio galaxy jets \citep{enslin97,ensslin98,mcnamara07}. They are visible in clusters in radio emission, %(CRe, and indirectly CRp which create secondary CRe through hadronic interactions),
and gamma-ray emission (via hadronic interactions). However, unlike in our interstellar medium (ISM), CRs in the ICM are energetically subdominant; for instance, current upper limits on CR-induced gamma-ray emission in Perseus suggest CRs are $\lsim 1-2\%$ of the thermal energy density \citep{aleksic12}. Why then are CRs in clusters of astrophysical significance? Firstly, unlike in the ISM, cosmic ray protons (CRp) with $E \lsim 10^{7}$GeV remain confined and have lifetimes of order a Hubble time \citep{volk96,berezinsky97}; they therefore encode archaeological information about the cluster assembly history as well as AGN and supernova activity. Secondly, the ICM provides stringent tests of plasma physics in a regime very different from the ISM. CRs in clusters represent an opportunity to study the unknown efficiency of shock acceleration \citep{blandford87} in a low Mach number ${\cal M} \sim 1-5$ and high plasma beta $\beta \sim 100$ regime. Transient radio phenomena can also teach us about magnetic field amplification at shocks. Thirdly, even a low level of CRs could have interesting astrophysical implications. These range from pressure support (thus affecting the use of clusters for cosmology) to heating which suppresses cooling flows \citep{Guo08a} or energizes filaments \citep{ferland08,ferland09}, and distributing metals and heat via buoyancy-induced turbulent convection \citep{chandran07,sharma09}.

% para on streaming speeds, application to radio halos, introduce Ensslin idea
% COMPARE WITH SUMMARY OF STREAMING IN EVERETT & ZWEIBEL
% Mention the Guo & Oh solution of have disrupted bubbles distribute CRs
One drawback of cosmological simulations of CRs in clusters is that they generally do not include CR transport processes; the CRs are assumed to be frozen into the gas, and advected with it. In practice, CRs can move relative to the gas by streaming along magnetic field lines down a CR gradient, as well as diffusing across field lines by scattering off plasma waves. As CRs stream, their momentum anisotropy excites plasma waves, which in turn scatter the CRs, isotropizing the CR distribution in the frame of the waves. This generally limits streaming speeds to the speed of the waves, which is the Alfv\'en speed $v_{\rm A}$. In our ISM, rapid pitch-angle scattering due to the CR streaming instability\footnote{In principle, CRs can also scatter off MHD turbulence, though this is thought to be weak due to the increasing anisotropy at small scales, with power concentrated in modes with wave-vectors transverse to the B-field, while CRs efficiently scatter off the parallel component \citep{chandran00,yan02}. Fast magnetosonic modes could potentially scatter CRs more efficiently \citep{brunetti07}, but a treatment of this is beyond the scope of this paper.} \citep{lerche67, kulsrud69, wentzel69, skilling71} can explain the observed spatial isotropy of CRs, as well as the escape time of CRs from the Galaxy \citep{schlickeiser02,kulsrud05}. Applying the same CR self-confinement scenario to the ICM, the low implied drift speed of CRs $v_{\rm D} \sim v_{\rm A} \sim 100 \, {\rm km \, s^{-1}}$, seems to justify neglect of cosmic ray transport. Early calculations of CRs in isolated clusters \citep{boehringer88,loewenstein91} which considered CR diffusion at a level comparable to ISM values found it to be negligible as a transport process. They argued that if CRs were injected at the cluster center by an AGN, they would quickly dominate pressure support at a level inconsistent with observations. \citet{Guo08a} resolved this in their calculations of CR heating by allowing CRs to be transported by rising buoyant bubbles, as seen in high resolution Chandra images, which are subsequently shredded by Kelvin-Helmholz and Rayleigh-Taylor instabilities to disperse the CRs.

However, this assumption of slow CR streaming and diffusion may not be fully justified. The plasma waves which scatter the CRs are also subject to a variety of damping mechanisms. If damping is stronger in the ICM than in the ISM, then pitch-angle scattering of the CRs off the attenuated waves will be reduced, and the CRs retain some momentum anisotropy in the frame of the waves. They can therefore stream faster than the waves, and will no longer be limited by the Alfv\'en speed. In principle, if the waves are very strongly damped, the CRs could stream up at speeds up to $\sim c$. While the possibility of super-Alfv\'enic or even free streaming was appreciated early on \citep{kulsrud69,kulsrud71,skilling71}, generally $v_{\rm D} \sim v_{\rm A}$ and diffusion coefficients appropriate for the ISM have been uncritically applied to the ICM environment. In an influential recent paper, \citet{enslin11} noted the interesting possibility of super-Alfv\'enic streaming in the ICM, adopted the sound speed $c_{\rm s}$ as a characteristic streaming speed, and were the first to discuss the wide-ranging observational consequences.

A particularly interesting possibility they focused on was whether the interplay between advection and streaming could be responsible for the observed bimodality seen in radio halo luminosity. Giant radio halos are generally only seen in disturbed clusters which show signs of merger activity.
% CONSIDER DISTINCTION BETWEEN GIANT HALO AND MINI-HALO---no turbulence in the latter...
This bimodality has been a stumbling block for hadronic models (e.g., \citet{pfrommer04}). These models track the long-lived CR protons (CRp) formed during structure formation shocks, and find that the secondary electrons formed when the CRps undergo hadronic interactions are sufficient to explain radio halo observations. Transience or correlation with turbulence is generally not expected in such models. As a result, radio halos are often attributed to the re-energization of seed electrons by Fermi II acceleration when the ICM turbulence becomes transonic during mergers \citep{brunetti01, brunetti07}. As the turbulence dies away, the CR electrons (CRe) cool via synchrotron and inverse Compton emission on a relatively short ($\sim 10^{8}$ yr) timescale. However, the origin of the seed electrons is uncertain; low energy electrons will rapidly Coulomb cool in the dense cluster center. \citet{enslin11} suggested instead that transonic turbulence advects CRs from the plentiful reservoir on the cluster outskirts. Hadronic interactions of the inwardly advected CRp with the dense central ICM can then produce CRe\footnote{Alternatively, low energy relic CRe advected from the cluster outskirts could provide seeds for turbulent reacceleration.}. Once turbulence dies down, subsequent outward CR streaming switches off the radio halo. This rapid outward streaming also explains why radio halos turn off in the original hadronic scenario, which is otherwise difficult to understand. For these explanations to work, the CR streaming timescale must be relatively short, or $v_{\rm D} \gg v_{\rm A}$. \citet{enslin11} adopted $v_{\rm D} \sim c_{\rm s}$ and examined its implications, but only justified this assumption qualitatively.

% goals of paper
This paper aims to critically examine the possibility that super-Alfv\'enic streaming could play a crucial role in CR transport in the ICM, by building more quantitative models to clarify its plausibility and importance.  It has three main goals: (i) a quantitative calculation of CR streaming speeds in quasi-linear theory and its dependence on plasma parameters, when a variety of wave damping mechanisms are at play. In particular, we consider the effects of non-linear Landau damping and turbulent damping. We also give expressions for parallel diffusivities and CR heating rates in this regime. Of particular interest is the countervailing effects of turbulence: it affects CR transport both by turbulent advection of CRs from the cluster outskirts, as well as damping of CR generated waves, which enables faster outward streaming. We assess their relative importance. (ii) A reevaluation of CR heating due to central injection by an AGN \citep{Guo08a}, taking these streaming effects into account. (iii) A 1D simulation of radio halo turnoff due to CR streaming, to establish if the required dimming by at least an order of magnitude can take place within a reasonable timescale. We also calculate how the gamma-ray luminosity evolves with time. A non-linear, time-dependent calculation is needed since the streaming speed itself depends on CR energy density. In their analytic calculations, \citet{enslin11} consider a steady-state profile where inward turbulent advection and outward CR streaming are in rough balance. This scenario seems somewhat unlikely; it seems more probable that at a given point in time, either inward advection or outward streaming dominates. To calculate radio halo turnoff, we consider the latter. These calculations also enable us to compute a fundamental prediction of this model: spectral steepening and frequency-dependent dimming which arise from energy-dependent streaming speeds. We compare these with observations.

% paper outline
The outline of this paper is as follows. In \S\ref{section:quasi-linear}, we calculate in quasi-linear theory cosmic-ray streaming speeds when different wave damping mechanisms are dominant, and derive expressions for the resulting parallel diffusivity, as well as turbulent diffusion rates. In \S\ref{sect:method}, we describe the equations we solve numerically with ZEUS,
% SPO: I took away comment on MHD, because we never use MHD capabilities
focusing in particular on the CR transport equation. We describe our initial conditions for the cosmic ray profile, which are tuned to match radio halo observations for the Perseus and Coma clusters. We also present a test case of CR heating by a central AGN. In \S\ref{results}, we present the results of simulations of radio halo turnoff due to CR streaming. In \S\ref{sect:analytic}, we show how aspects of these results can be understood analytically.  Finally, we conclude in \S\ref{sect:conclusions}.

%% ADD IN APPENDIX IF DECIDE TO DO THAT

\section{Cosmic Ray Streaming: Quasi-linear Theory}
\label{section:quasi-linear}

In this section, we derive the basic equations we use, in particular the streaming speeds and diffusion coefficients which are used in the cosmic-ray transport equation. Our treatment is by design semi-quantitive rather than fully rigorous.

\subsection{Cosmic Ray Streaming}\label{section:stream}

\subsubsection{Resonant Scattering and Wave Growth}
\label{section:growth}

Consider a cosmic ray proton with Lorentz factor $\gamma$ propagating along a magnetic field line of strength $B_0$ with cyclotron frequency $\Omega_0 = eB_0/(m_pc)$, gyroradius $r_\tr{L} = \gamma c/\Omega_0$, and pitch angle cosine $\mu$. Since $v_{\rm A} \ll c$, Alfv\'en waves are perceived by the CR as a spatially varying but time-stationary B-field. An Alfv\'en wave is resonant with this cosmic ray if it has a wave vector $\mathbf{k}$ whose component parallel to the magnetic field $k_{\parallel}$ equals the parallel projection of the gyroradius, i.e. if the resonance condition
\begin{equation}
k_{\parallel} = \frac{1}{\mu r_\tr{L}}
\label{eqn:resonance}
\end{equation}
is satisfied. This resonance is a requirement both for the wave to scatter the CR and for the CR to excite the wave. This condition can be easily understood: if the magnetic field changes on a length scale much longer than the projected gyroradius, the CR will simply follow the field line adiabatically, with no change in pitch angle. If the field varies on much smaller scales, the CR will see a rapidly oscillating Lorentz force during its orbit and remain unaffected, essentially only seeing the background field. At resonance, the CR sees a constant field due to the wave, and hence a steady force. The $k_{\parallel}$ portion of the wave is the relevant one, since it has a transverse magnetic field $\delta B_{\perp}$ which can exert a Lorentz force on the $v_{\parallel}$ component of a streaming cosmic-ray.

If the distribution of cosmic rays in the frame of the Alfv\'en waves is completely isotropic, then the effect of a cosmic ray traveling along the magnetic field line in one direction is cancelled by an equivalent cosmic ray traveling in the opposite direction, and there is no wave growth. However, \cite{kulsrud69} showed that even a slight anisotropy in the cosmic rays--which naturally arises in the presence of sources and sinks--will cause unstable growth in the waves, caused by momentum transfer from the CRs to the waves in the course of pitch-angle scattering. The resulting wave growth rate is \citep{kulsrud69}:
\begin{equation}\label{eqn:growth}
\Gamma_{\tr{CR}}(k_{\parallel}) \sim \Omega_0 \frac{n_\tr{CR}(>\gamma)}{n_\tr{i}}\left(\frac{v_\tr{D}}{v_\tr{A}}-1\right)
\end{equation}
In the above, $n_\tr{CR}(>\gamma)$ is the number density of cosmic ray protons with energies large enough to be resonant with the Alfv\'en wave for some pitch angle $\mu$, namely $r_\tr{L} > 1/k_{\parallel}$ (though since the CR spectrum falls off rapidly with energy, generally $k_{\parallel} \sim 1/r_{\rm L}$), $n_\tr{i}$ is the ion density in the plasma, and $v_\tr{D}$ and $v_\tr{A}$ are the cosmic ray streaming and Alfv\'en speeds respectively. This expression is derived from balancing CR momentum loss with wave momentum gain, but we can understand its main features qualitatively. The rate of wave growth scales with that for momentum loss for a single CR, $\dot{p} \propto p \Omega_{\rm rel} = (\gamma m c^{2}) (\Omega_{\rm o}/\gamma) \propto \Omega_{o}$, i.e. the non-relativistic, rather than the relativistic gyro-frequency. This has to be multiplied by the fraction of ions which can drive wave growth, $n_\tr{CR}(>\gamma)/n_\tr{i}$ (non-resonant ions simply provide inertia, slowing down wave growth), and the anisotropy which seeds the wave growth $(v_{\rm D}/v_{\rm A}-1)$.

The streaming instability causes the waves to grow until pitch-angle scattering renders the CR distribution isotropic in the frame of the waves, i.e. $v_{\rm D} \sim v_{\rm A}$. If we assume $(v_{\rm D}/v_{\rm A} -1) \sim \cal{O}$(1), we can estimate the growth time of the waves. If we assume that the energy density is CRs is $\sim 10\%$ of the thermal energy density, $\epsilon_{\rm CR} \sim 0.1 \epsilon_{\rm therm}$, then $n_{\rm CR} \langle E_{\rm CR} \rangle \sim 0.1  n_{\rm i} \langle E_{\rm i} \rangle$ where $\langle E_{\rm CR} \rangle \sim$ GeV (as is true for most reasonable power-law momentum distributions--e.g., see Fig. 1 of \citet{enslin07}), and $\langle E_{\rm i} \rangle \sim$ keV are the typical energies of CRs  and thermal ions respectively, {and so} $n_{\rm CR}/n_{\rm i} \sim 10^{-7}$. A similar ratio holds in the coronal regions of our Galaxy. For $\sim \mu$G fields, $\Omega_{\rm o} = eB/mc \sim 10^{-2} \, {\rm s^{-1}}$, implying from equation (\ref{eqn:growth}) a wave growth time of $\Gamma_{\tr{CR}}^{-1} \sim 30$ yr, i.e. extremely short. %The required perturbations to the B-field for self-confinement are extremely small. One can show that for a single CR-wave interaction, $\delta \theta \sim \delta B/B$ (e.g., \citet{kulsrud05}). Thus, the variance of the pitch angle $\langle (\delta \theta)^{2} \rangle$ will random walk at a rate ${\rm D}_{\theta} \sim \Omega (\delta B/B)^{2}$. The lifetime of CRs in the disk of our Galaxy (with scale height $L \sim 3$ kpc) from spallation measurements is $t \sim 3 \times 10^{6}$yr, which implies a mean free path of $\lambda \sim L^{2}/ct \sim 10$pc. For

The above arguments suggest that self-confinement of cosmic-rays is very efficient, and should always reduce the streaming velocities $v_{\rm D} \sim v_{\rm A}$. The general success of the self-confinement picture for our Galaxy means that this has been uncritically assumed in other environments such as the ICM, and/or CR diffusion coefficients scaled to the measured Galactic values. In fact, $v_{\rm D}$, and the associated diffusion coefficient $D_{\theta}$ depend on the amplitude of the wave field $\delta B/B$, which can be calculated by assuming equilibrium between growth and damping. If damping processes are sufficiently strong, then $\delta B/B$ will be insufficient to efficiently confine the CRs, and super-Alfv\'enic streaming is possible. We now examine this possibility.

\subsubsection{Non-linear Landau Damping}
\label{section:landau}

Parallel propagating MHD waves do not suffer any linear damping. However, they can undergo non-linear Landau damping when two waves A \& B of slightly different frequency interact to form a beat wave. This beat wave can resonantly interact with thermal particles with parallel velocity identical to the wave's phase speed, $v_{\parallel} = (\omega_{\rm A}-\omega_{\rm B})/(k_{\rm A}-k_{\rm B})$ (for parallel propagating waves, $v_{\parallel} = v_{\rm A}$). Particles moving more slowly than the beat wave will extract energy from the wave (thus damping it), while particles moving faster than the wave will add energy to it. For a Maxwellian plasma, typically $(\partial f/\partial v)_{v_{\parallel}=v_{\rm A}} < 0$, and damping dominates.\footnote{In a high $\beta$ plasma,  $(\partial f/\partial v)_{v_{\parallel}=v_{\rm A}}$ is relatively flat for electrons,  while still steep for ions; hence, ions dominate the damping rate \citep{miller91}.} The high frequency wave gives up energy to a combination of the low frequency wave and resonant particles.

To aid in physical insight, we present a simplified derivation of streaming speeds to be expected if non-linear Landau damping dominates, before employing the formulae from more detailed derivations \citep{lee73,kulsrud78,felice01}. The damping rate in a high-$\beta$ plasma is \citep{kulsrud05}:
\begin{equation}
\Gamma_{\rm NL} \approx \frac{1}{2}\sqrt{\frac{\pi}{2}} \left( \frac{v_{\rm i}}{v_{\rm A}} \right) \left( \frac{\delta B}{B} \right)^{2} \omega \approx 0.3 \frac{\Omega}{\mu} \frac{v_{\rm i}}{c} \left( \frac{\delta B}{B} \right)^{2}
\label{eqn:NLL_damp}
\end{equation}
How can we qualitatively understand the first relation? The wave frequency $\omega = k_{\parallel} v_{\rm A}$ sets the fundamental frequency, while interactions involving the beat wave arise to second-order in perturbed field strength $(\delta B/B)^{2}$. Since the resonant condition $v_{\parallel}=\mu v_{\rm i} = v_{\rm A}$ implies that thermal particles with $\mu = v_{\rm A}/v_{\rm i}$ are resonant, it is clear that the damping rate should depend on this ratio. However, the exact dependence only emerges from a detailed calculation--either by calculating the slope $(\partial f/\partial v)_{v_{\parallel}=v_{\rm A}}$, or from the plasma dispersion relation \citep{foote79}. Note that since  $v_{\rm i}/v_{\rm A} = \beta^{1/2}/2$, we have $\Gamma_{\rm NL} \propto \beta^{1/2}$. In the second relation, we use the dispersion relation $\omega = k_{\parallel} v_{\rm A}$ and the resonance condition, equation (\ref{eqn:resonance}).

In steady state, the Vlasov equation for CRs is \citep{kulsrud69}:
\begin{equation}\label{vlasov1}
v_{\rm z} \frac{\partial f}{\partial z} = \frac{\partial}{\partial \mu} \left[ \frac{1 - \mu^{2}}{2} \nu(\mu) \frac{\partial f}{\partial \mu} \right]
\end{equation}
where $B=B_{z} \hat{z}$, $\nu (\mu) \approx \Omega (\delta B/B)^{2}$ and $v_{\rm z} = \mu c$.
This expresses the condition that the net streaming along field lines is set by diffusion in pitch-angle. In the limit of strong scattering, we can expand $f=f_{0} + f_{1} + f_{2} + ...$ where $f_{0}(p,z,t)$ is isotropic and $f_{1}(p,z,t,\mu) \ll f_{0}$, $f_{2} \ll f_{1}$. Let us also define ${\cal F} = f_{1}/f_{0}$ and the scale height $L_{z}(p,z) = f_{0}/(\partial f/\partial z)$. Integrating both sides of equation \eqref{vlasov1} with respect to $\mu$ and dividing by $f_0$, we obtain:
\begin{equation}
\frac{\partial {\cal F}}{\partial \mu} = - \frac{c}{\nu L_{z}}.
\end{equation}
If we set
\begin{equation}
{\cal F} = 1 + \frac{3 (v_{\rm D} - v_{\rm A})}{c} \mu
\end{equation}
so that $\langle \mu c {\cal F}(\mu) \rangle = (v_{\rm D} - v_{\rm A})$ (i.e., the leading order anisotropy in the distribution function yields the net drift relative to the frame of the waves), this yields:
\begin{equation}
(v_{\rm D} - v_{\rm A}) \approx  \frac{r_{\rm L}}{3 L_{\rm z}} \left( \frac{\delta B}{B} \right)^{-2} c \approx \frac{\lambda}{3 L_{\rm z}} c
\label{eqn:v_D_intermediate}
\end{equation}
where $\lambda \sim r_{\rm L} (\delta B/B)^{-2}$ is the mean free path. In steady state, the wave growth rate (equation \eqref{eqn:growth}) equals the wave damping rate (equation \eqref{eqn:NLL_damp}). Together with equation (\ref{eqn:v_D_intermediate}), this gives us two equations which we can solve for two unknowns, $v_{\rm D}$ and $(\delta B/B)^{2}$. The result is:

\begin{equation}
\left(\frac{\delta B}{B} \right)^{2} = %\left( \frac{2}{\pi} \right)^{1/4} \left( \frac{n-3}{n-2} \right)^{1/2}
\left( \frac{c}{v_{\rm A}} \frac{c}{v_{\rm i}} \frac{r_{0}}{3 L_{\rm z}} \frac{n_{\rm CR}(> \gamma)}{n_{i}} \gamma^{2} \right)^{1/2}
\end{equation}
where $f_{0} \propto p^{-n}$ (note that our result differs from \citet{felice01}, who explicitly specialize to $\gamma \sim 5$ for the ISM from the outset). If we scale to numbers characteristic of the ICM, we obtain:
\begin{equation}
\left(\frac{\delta B}{B} \right)^{2} = 1.6 \times 10^{-6} \frac{ (n^{\rm CR}_{-10})^{1/2} \gamma_{100}^{(5-n)/2} 10^{4.6-n}} {(n_{-3}^{i})^{1/4}B_{\rm \mu G} T_{\rm 4 \, keV}^{1/4} L_{\rm z,100}^{1/2}}
\label{eqn:Bsq_NLLD}
\end{equation}
where $T_{\rm 4 \, keV}=(T/4 \, {\rm keV})$, $B_{\rm \mu G}=(B/1 \, \mu G)$, $L_{\rm z,100}=(L_{z}/100 \, {\rm kpc})$, $n^{\rm i}_{-3}=(n_{\rm i}/10^{-3} \, {\rm cm^{-3}})$, $n^{\rm CR}_{-10}=n^{CR}(\gamma > 1)/10^{-10} \, {\rm cm^{-3}})$, $\gamma_{100}=\gamma/100$, and we have scaled to $n=4.6$. Note that $n^{CR} (> \gamma) = 10^{-10} \gamma^{-1.6} \, {\rm cm^{-3}}$ roughly corresponds to a CR energy density in equipartition with a $\sim \mu$G B-field. The fact that $(\delta B/B)^{2} \ll 1$ self-consistently implies that quasi-linear theory is applicable. If we insert this into equation (\ref{eqn:v_D_intermediate}), we obtain for the drift speed:
\begin{equation}
v_{\rm D} = v_{\rm A} \left( 1+  0.9 \frac{(n_{-3}^{i})^{3/4} T_{\rm 4 \, keV}^{1/4}  10^{n-4.6}}{B_{\rm \mu G}L_{\rm z,100}^{1/2} (n^{\rm CR}_{-10})^{1/2} }  \gamma_{100}^{(n-3)/2} \right)
\label{eqn:vD_NLLD}
\end{equation}

Several points should be noted. For these parameters, streaming speeds do not significantly exceed the Alfv\'en speed for the $\sim$100 GeV cosmic-ray protons which in turn produce the 10 GeV CR electrons which in turn produce $\sim$GHz radio emission. For Alfv\'en speeds of $v_{\rm A} \approx 70 \, {\rm km \, s^{-1}} \, B_{\rm \mu G} n^{-1/2}_{i,-3}$, this implies radio halo turnoff times of $t \sim 1.4 \, {\rm Gyr} \, L_{100} B_{\rm \mu G}^{-1} n^{1/2}_{i,-3}$, which may seem too long. However, note that $L_{\rm z}, n^{\rm CR}$ will be time-dependent functions during the streaming process, so it is necessary to check how streaming evolves in a time-dependent calculation. Our results should be contrasted with those of \citet{enslin11}, who describe similar estimates based on \citet{felice01}, but do not give explicit expressions. Unlike them, we find $v_{\rm D} \ll c_{\rm s}$ for plasma parameters corresponding to observed clusters; nothing in the problem singles out the sound speed as a reference speed. Note that all the parameters in equation (\ref{eqn:vD_NLLD}) are observationally constrained, so order of magnitude departures are unlikely. Also note that even though $\Gamma_{\rm NL} \propto \beta^{1/2}$, there is no explicit $\beta$ dependence in $v_{\rm D}$.

\subsubsection{Turbulent Damping}
\label{section:turbulence}

Another source of wave damping comes from the highly anisotropic nature of MHD turbulence \citep{farmer04,yan02}.
We adopt the \citet{goldreich95} theory (hereafter 'GS') for strong, incompressible MHD turbulence; an excellent summary can be found in \citet{lithwick01}. Turbulence in clusters is generally incompressible since it is significantly subsonic except at the cluster periphery. The strong turbulence regime where GS theory is applicable sets in at wavenumbers $k \lsim k_{o} M_{\rm A}^{-2}$ (e.g., see \citet{nazarenko11}); since $M_{\rm A} \gsim 1$ in clusters, the theory is clearly applicable, particularly at the small scales relevant for CR scattering. GS theory has support both from numerical simulations \citep{cho00,maron01} and solar wind measurements \citep{horbury08,podesta09,wicks10,chen11}. It is anisotropic, with:
\begin{eqnarray}
\label{eqn:GS1}
v_{\lambda_{\perp}} &\sim& v_{\rm A} \left( \frac{\lambda_{\perp}}{L_{\rm MHD}}\right)^{1/3} \sim (\epsilon \lambda_{\perp})^{1/3}
 \\
\frac{\Lambda_{\parallel}(\lambda_{\perp})}{\lambda_{\perp}}  &\sim& \left( \frac{L_{\rm MHD}}{\lambda_{\perp}}\right)^{1/3}
\label{eqn:GS2}
\end{eqnarray}
where $\lambda_{\perp}$ is the length scale transverse to the local mean magnetic field, $v_{\lambda_{\perp}}$ is the rms velocity fluctuation across $\lambda_{\perp}$, $\Lambda_{\parallel}(\lambda_{\perp})$ is the length scale parallel to the local mean magnetic field across which the velocity fluctuation is $v_{\lambda_{\perp}}$,$L_{\rm MHD}$ is the length scale at which turbulence is excited with velocity perturbations comparable to the Alfv\'en speed $v_{\rm A}$ (i.e., with $M_{\rm A} \sim 1$), and $\epsilon \sim v_{\lambda_{\perp}}^{3}/\lambda_{\perp} \sim v_{\rm A}^{3}/L_{\rm MHD}$ is the (constant) energy cascade rate per unit mass. Note that $L_{\rm MHD}$ is {\it defined} to be the scale at which $M_{\rm A}=1$; if turbulence is already sub-Alfv\'enic at the outer scale, then it should be considered an extrapolation. Equation (\ref{eqn:GS1}) describes a standard Kolmogorov cascade in the transverse direction. Equation (\ref{eqn:GS2}) indicates that an eddy becomes increasingly elongated along the magnetic field, $\Lambda_{\parallel} \gg \lambda_{\perp}$ as the cascade proceeds deep into the inertial range $\lambda_{\perp} \ll L_{\rm MHD}$. It can be derived from the assumption of ``critical balance'', which states that characteristic linear and non-linear interaction times are approximately equal at all scales (e.g., see \citet{nazarenko11}). Thus, the cascade proceeds primarily in the transverse direction, and most of the power is concentrated in modes with transverse wave vectors. Intuitively, we can understand this from the fact that in MHD turbulence, non-linear interactions arise from collisions of oppositely directed Alfv\'en wave packets travelling along field lines. A wave packet is distorted when it follows field lines perturbed by its collision partner; it cascades when the field lines along which it is propagating have spread by a distance comparable to $\lambda_{\perp}$. Since the magnetic and velocity fluctuations associated with Alfv\'en waves are transverse to the local mean field, the cascade proceeds primarily in the transverse direction.

Turbulence therefore suppresses the waves responsible for self-confinement of cosmic rays, since they cascade to smaller scales before they have an opportunity to scatter CRs. In particular, the small scale transverse components injected by the cascade mean that the CR no longer experiences a time-steady force in its orbit; instead it sees an oscillating force which leads to inefficient scattering. For these same reasons, MHD turbulence scatters CRs inefficiently \citep{chandran00,yan02}. The damping rate of a wave is simply the eddy turnover rate
\citep{farmer04}:
\begin{equation}
\Gamma_{\rm turb} \sim \frac{v_{\lambda_{\perp}}}{\lambda_{\perp}} \sim \frac{\epsilon^{1/3}}{\lambda_{\perp}^{2/3}}
\label{eqn:turbdamp1}
\end{equation}
where we use equation (\ref{eqn:GS1}). Growth rates are highest, and damping rates lowest, for the most closely parallel-propagating waves, i.e. those with the largest $\lambda_{\perp}$. Even if a CR-generated wave starts out as parallel-propagating, the turbulent cascade injects transverse components which subsequently cascade. The amplitude of magnetic field fluctuations across a scale $\lambda_{\perp}$ thus define a minimal aspect ratio\footnote{This should not be confused with the eddy aspect ratio $\Lambda_{\parallel}(\lambda_{\perp})/\lambda_{\perp} \gg 1$, whereas we have $(\lambda_{\parallel}/\lambda_{\perp})_{\rm min}  < 1$. Typical eddies in the MHD cascade vary mostly in the transverse direction, $k_{\perp} \gg k_{\parallel}$,  whereas we seek waves injected by CRs with the least possible transverse variation, $k_{\perp} \ll k_{\parallel}$.} $(k_{\perp}/k_{\parallel})_{\rm min} \sim \delta B(\lambda_{\perp})/B \sim v_{\lambda_{\perp}}/v_{\rm A} \sim (\lambda_{\perp}/L_{\rm MHD})^{1/3}$ (using equation \eqref{eqn:GS1} in the last step). From the resonance condition $k_{\parallel}^{-1} \sim r_{\rm L}$, the smallest possible perpendicular wavenumber is $k_{\perp,{\rm min}} \sim \epsilon^{1/4}(r_{\rm L} v_{\rm A})^{-3/4}$. Inserting the largest possible perpendicular wavelength $\lambda_{\perp} \sim k_{\perp,{\rm min}}$ into equation (\ref{eqn:turbdamp1}), the minimal damping rate for a wave with $k_{\parallel} \sim r_{\rm L}^{-1}$ is \citep{farmer04}:
\begin{equation}
\Gamma_{\rm turb,min} \sim \left( \frac{\epsilon}{r_{\rm L}v_{\rm A}} \right)^{1/2}.
\label{eqn:gamma_turb}
\end{equation}
If in steady state we balance the wave growth rate (equation \eqref{eqn:growth}) with this damping rate, we obtain a streaming speed:
\begin{equation}
v_{\rm D} = v_{\rm A} \left( 1+ 1.2 \frac{B_{\rm \mu G}^{1/2} n_{i,-3}^{1/2}}{L_{\rm MHD, 100}^{1/2} n_{\rm CR,-10}} \gamma_{100}^{n-3.5} 10^{{2(n-4.6)}}\right)
\label{eqn:vD_turb}
\end{equation}
where $L_{\rm MHD, 100}=L_{\rm MHD}/100$ kpc. We also obtain $(\delta B/B) \sim 10^{-3}$. At first blush, non-linear Landau damping and turbulent damping both seem to give similarly slow streaming speeds. However, note that $v_{\rm D} - v_{\rm A} \propto (n_{\rm CR}^{-1/2},n_{\rm CR}^{-1})$ for these two sources of damping respectively. This difference becomes crucial during non-linear evolution, enabling CRs in the turbulent damping case to stream much more effectively.

\subsubsection{General Remarks on Cosmic-Ray Streaming}
\label{section:stream_conclude}

We have now derived streaming speeds for two different damping mechanisms, which depend both on CR energy ($v_{\rm D} \propto \gamma^{0.8}, \gamma^{1.1}$ for non-linear Landau damping and turbulent damping respectively) and plasma parameters -- most notably the CR number density. The streaming speed is thus a function of both position and time, and is best self-consistently solved in a time-dependent calculation, as we will soon tackle.\footnote{We have also assume $n(>\gamma)$ to be a fixed power-law, whereas it steepens with time due to energy dependent streaming.} Before we forge ahead and use these expressions, there are several potential complications worth discussing.

Our streaming speeds for the ICM are characteristically of order the Alfv\'en speed, although this can vary spatially and temporally as plasma parameters vary, particularly the CR number density. \citet{enslin11} argue against the Alfv\'en speed as a characteristic CR propagation speed in a high $\beta$ plasma, arguing that in the limit where the background magnetic field $B \rightarrow 0$, this would imply that $v_{\rm D} \approx v_{\rm A} \rightarrow 0$, rather than $v_{\rm D} \rightarrow c$, as might be expected if there is no magnetic field to couple the CRs to the plasma. Instead, they advocate the sound speed $c_{\rm S}$ as a characteristic streaming speed. We have several remarks. The streaming speeds we have calculated via quasi-linear theory assumes $(\delta B/B) \ll 1$, and we have checked that this condition is self-consistently fulfilled in the ICM (typically, $(\delta B/B) \sim 10^{-4}$), an amplitude similar to that inferred for the coronal gas in our Galaxy. The hypothetical limit $B \rightarrow 0$ (which is not realized in the ICM) clearly violates this assumption, and requires a fully non-linear calculation. There, we might expect that instabilities generated by a current of streaming CRs (e.g., \citet{bell78})  would nonetheless generate a B-field which will confine the CRs. Nothing in our calculations singles out the sound speed as a reference velocity.

The resonance condition, equation (\ref{eqn:resonance}), shows that CRs of larger pitch angle ($\mu \rightarrow 0$) interact with waves of progressively shorter wavelength. However, growth rates $\Gamma_{\rm CR} \propto \mu$ (e.g., \citet{kulsrud05}), while non-linear Landau damping $\Gamma_{\rm NL} \propto 1/\mu$ (equation \eqref{eqn:NLL_damp}), so there is relatively little energy in such short wavelength waves as $\mu \rightarrow 0$. On the face of it, this would imply that it is impossible for particles to scatter across the $\theta=90^{\circ}$ point via resonant scattering to reverse direction, the well-known `$90^{\circ}$ problem' (in fact, the affected region is small; quasi-linear interactions can effectively scatter CRs down to $\mu_{\rm c} \sim 10^{-4}$. The gap is  a little larger, $\sim v_{i}/c \sim 3 \times 10^{-3}$, in a high-$\beta$ plasma when ion-cyclotron damping is effective \citep{holman79}). The fact that CRs appear to be efficiently confined and isotropized in our Galaxy implies that Nature has found a way around it. The leading explanation appears to be mirror interactions from MHD waves created by the $\theta \sim 0^{\circ}$ CRs, which are able to trap the particles and turn them around \citep{felice01}. These mirror interactions can also be thought of as resonance broadening \citep{achterberg81,yan08} of the long wavelength waves. \citet{felice01} conduct a detailed boundary layer calculation of the mirror interaction and find that it introduces a minor logarithmic correction (which we have ignored) to the standard calculation. We note that if there were indeed a $90^{\circ}$ problem in the ICM, the resulting light-speed streaming speeds would imply flat CRp and CRe profiles, which is inconsistent at least with observations of radio mini-halos such as Perseus. It would also shut off radio halos extremely rapidly, regardless of how the relativistic electrons are produced.

The wave damping rate is the sum of all damping processes, and thus in principle one should always consider the contribution from both turbulent and non-linear Landau damping. In practice, we consider limiting regimes where one process dominates. Their ratio is:
\begin{equation}
\frac{\Gamma_{\rm turb}}{\Gamma_{\rm NL}} \approx 1. \frac{B_{\rm \mu G}^{3/2} n_{i,-3}^{1/4} L_{\rm z,100}^{1/2}}{L_{\rm MHD,100}^{1/2}T_\tr{4keV}^{1/4} n_{\rm CR,-10}^{1/2}} \gamma_{100}^{n/2-2} \propto \left( \frac{1}{\nabla f} \right)^{1/2}.
\end{equation}
Turbulent damping thus always dominates at late times as the CR profile falls ($n_{\rm CR} \rightarrow 0$) and flattens ($L_{\rm z} \rightarrow \infty$).

We have only considered CR self-confinement, and ignored other possible mechanisms for scattering CRs. As we have previously discussed, the anisotropic nature nature of Alfv\'enic MHD turbulence (which is mostly transverse on small scales comparable to the gyro-radius, in contrast to the parallel modes required to scatter CRs) make them inefficient scatterers of CRs \citep{chandran00,yan04}. While the distribution of slow magnetosonic waves follows that of Alfv\'en waves \citep{lithwick01}, fast magnetosonic modes can potentially have an independent non-linear cascade which is isotropic and can efficiently scatter CRs \citep{schlickeiser02,brunetti07}. For now, we eschew this possibility, in favor of the well-established self-confinement picture, which is the generally accepted theory in our Galaxy. One failing of the self-confinement picture in our Galaxy is that both non-linear Landau damping and turbulent damping appear to damp the waves too efficiently at high energies; the increase of streaming speeds with energy appear inconsistent  with the low observed CR anistropy for $E > 100$ GeV \citep{farmer04}. \citet{chandran00} has proposed that magnetic mirror interactions in dense molecular clouds could provide this further confinement, though the possibility remains that some aspects of the physics are still not well understood. A conservative reading of these possible complications would take our derived streaming speeds and diffusion coefficients as upper bounds; they could potentially be lower if scattering is more efficient.

\subsection{Cosmic-Ray Transport}
\label{section:transport}

\subsubsection{Cosmic-Ray Transport Equation}
\label{section:transport-equation}

The cosmic ray transport equation in the limit of large wave-particle scattering is \citep{skilling71}:
\begin{equation}
\begin{split}
\frac{\partial  f_\tr{p}}{\partial t}+(\mathbf{u}+\mathbf{v}_\tr{A})\cdot\nabla  f_\tr{p}&=\nabla\cdot(\kappa_\tr{p} \mathbf{nn}\cdot\nabla  f_\tr{p})\\ &+\frac{1}{3}p\frac{\partial  f_\tr{p}}{\partial p} \nabla\cdot (\mathbf{u}+\mathbf{v}_\tr{A})+ Q
\label{eqn:crevol}
\end{split}
\end{equation}
Here, $ f_\tr{p}(\mathbf{x},p,t)$ is the cosmic ray distribution function (isotropic in momentum space), $u$ is the gas velocity, $\mathbf{v_\tr{A}}$ is the Alfv\'en velocity, $\mathbf{n}$ is a unit vector pointing along the magnetic field, and $ Q$ is a cosmic ray source function. Throughout this paper, we shall always use the 3D distribution function $ f_\tr{p}$, which does not include the differential volume factor $4\pi p^2$. All momenta $p$, unless otherwise specified, will always be in units of $mc$ throughout this paper. The actual momentum will be written \textbf{$\tilde{p}_i=p_im_ic$}, the subscript denoting the particle type. Any distribution functions written as functions of particle energy rather than momentum will be related by
\begin{equation}
\tr{d}n_i={4\pi p^2}f_i(p_i)\tr{d}p_i=f_i(E_i)\tr{d}E_i
\end{equation}
\[
E_i=\sqrt{1+p_i^2}m_ic^2\rightarrow\tr{d}E_i=\frac{p_im_ic^2\tr{d}p_i}{\sqrt{1+p_i^2}}
\]
Equation (\ref{eqn:crevol}) is derived from the collisionless Vlasov equation, which expresses conservation of phase space density:
\begin{equation}
\pp{ f}{t}+\nabla\cdot( f\mathbf{v})+\nabla_p\cdot\left( f\pp{\mathbf{p}}{t}\right)=0
\end{equation}
but evaluated in the frame of the Alfv\'en waves (which has velocity $\mathbf{u} + \mathbf{v}_\tr{A}$, the sum of the local gas and Alfv\'en velocities). The distribution function is then expanded in inverse powers of the CR-wave collision frequency $\nu$, $f=f_{0} + f_{1} + f_{2}+...$, where $f_{r} = {\cal O}(\nu^{-r})$. Equation (\ref{eqn:crevol}) is obtained after averaging over pitch angle (justified in the limit of frequent scattering), and is accurate to second order, ${\cal O}(\nu^{-2})$. The term with $\kappa_{p}$ expresses diffusion relative to the wave frame, and is discussed in detail below. For the details of this expansion we refer the reader to \cite{skilling71}. This equation implicitly assumes that $f_{0} \gg f_{1}$, i.e. to leading order strong wave-particle scattering renders the distribution function isotropic in the wave frame. As we have seen, for most plasma parameters $(v_{\rm D}-v_{\rm A})/c \ll 1$, so this assumption is justified.

The physical interpretation of equation (\ref{eqn:crevol}) is easy to understand. The left-hand side of this equation is a total time derivative, including an advection term in the frame of the waves. The first two terms on the right-hand side represent diffusion along magnetic field lines relative to the wave frame and adiabatic losses/gains respectively. As long as we have a functional form for $\kappa_\tr{p}$ and $ Q$, this equation completely describes the evolution of the cosmic ray population.

Note that in the frame of the wave, and considering the isotropic part of the distribution function $f_{0}$ (so that there is no diffusion relative to the wave frame, $\kappa_{p}=0$), we have:
\begin{equation}
\frac{D f_{0}}{D t} = \frac{1}{3} p \pp{f_{0}}{p} \nabla \cdot (\mathbf{u} + \mathbf{v}_\tr{A})
\end{equation}
i.e. the CRs evolve adiabatically in the wave frame, with $p \propto n_{\rm CR}^{1/3}$ \citep{skilling71}.
This makes physical sense: there are no electric fields in the frame of the wave, and hence the particles conserve energy; they can only scatter in pitch angle. However, the CRs do {\it not} evolve adiabatically in the frame of the gas, where there are electric fields associated with the hydromagnetic waves. Thus, there is an irreversible energy transfer from the CRs to the gas, with volumetric heating rate (e.g., \citet{kulsrud05}):
\begin{equation}
\Gamma_{\rm wave}= -  \mathbf{v}_\tr{A} \cdot \nabla P_{c},
\end{equation}
which we shall refer to as the ``wave heating rate''.
This may be thought of as the rate at which work is done on the gas by CR pressure forces, $\mathbf{v_{A}} \cdot \mathbf{F}$. Importantly, this heating rate is {\it not} $\Gamma_{\rm wave}= -  \mathbf{v}_\tr{D} \cdot \nabla P_{c}$, as has sometimes been adopted elsewhere in the literature (e.g., \citet{uhlig12}). The latter expression gives rise to unphysically large heating rates when $v_{\rm D} \gg v_{\rm A}$. Super-Alfv\'enic streaming arises due to a {\it reduction} in coupling between CRs and gas; it is unphysical that this would give rise to greater heating. Physically, all momentum and energy transfer between the CRs and gas is mediated by hydromagnetic waves; the rate at which work is done by any transmitted forces is therefore set by the velocity of the waves $\mathbf{v}_{\rm A}$.

To next order in $\nu^{-1}$, slippage with respect to the wave frame is expressed by the diffusion coefficient $\kappa_{p}$:
\begin{equation}\label{eqn:kappap}
\kappa(\gamma)= c^2 \Big\langle\frac{1-\mu^2}{\nu(\mu,\gamma)}\Big\rangle
\end{equation}
where the wave-particle collision frequency $\nu(\mu,\gamma)$ is \citep{kulsrud69}:
\begin{equation}
\nu(\mu,\gamma) = \frac{\pi}{4} \Omega_{0} \left( \frac{\delta B}{B} \right)^{2} (\mu,\gamma)
\label{eqn:nu_collide}
\end{equation}
and the average is taken over pitch angle \citep{skilling71}. This expression is obtained from equation (\ref{vlasov1}) as shown by \citet{kulsrud69}, and we assume relativistic CRs such that $v \sim c$. From equation (\ref{eqn:kappap}), the more frequently CRs interact with Alfv\'en waves, the more slowly they diffuse relative to the waves--- as one might expect, since scattering isotropizes the CR in the wave frame. Equation (\ref{eqn:nu_collide}) can be understood from the fact that a single CR-wave encounter in one gyro-period $\tau$ leads to a change in pitch angle $\Delta \theta \approx (\delta B/B)$ \citep{kulsrud05}; thus $N \sim t/\tau$ encounters leads to a net random walk in pitch angle of $(\Delta \theta)^2  \sim N (\delta B/B)^{2} \sim t/\tau (\delta B/B)^{2}$, or a pitch angle diffusion rate of $D_{\theta} \sim (\Delta \theta)^2/t \sim \Omega_{0} (\delta B/B)^{2}$\footnote{The mean free path of a CR is roughly the distance over which the pitch angle diffuses by order unity (so that the CR reverses direction), $\lambda \sim c D_{\theta}^{-1} \sim 3 \times 10^{12} (\delta B/B)^{-2}$ cm, where the pitch angle diffusion coefficient $D_{\theta} \sim (\delta B/B)^{-2} \Omega_{0}$. Thus, even small fields of ($\delta B/B) \sim 10^{-3}$ would lead to mean free paths of $\lambda \sim 1$ pc, implying that the diffusive approximation is excellent.}. Equation (\ref{eqn:kappap}) can be evaluated by equating wave growth and damping rates to obtain the amplitude of the waves, $(\delta B/B)^{2}$, as for instance in equation (\ref{eqn:Bsq_NLLD}). It can also be intuitively written in terms of streaming speeds. From equation (\ref{eqn:crevol}), we can write the net streaming speed (i.e. the frame in which the mean CR flux vanishes), as \citep{blandford87}:
\begin{equation}\label{vDrelation}
v_\tr{D}=\frac{1}{ f_\tr{p}(p)}\left[-\frac{1}{3}v_\tr{A}p\frac{\partial  f_\tr{p}}{\partial p}-\kappa \mathbf{n}\cdot\nabla  f_\tr{p}\right]
\end{equation}
where ${\mathbf n}$
is a unit vector pointing along the magnetic field, down the CR gradient. The first term effectively corrects for the Compton-Getting effect, i.e. the differential Doppler shifts of particle energies in transforming from the wave to the inertial frame (depending on whether particles are moving parallel or anti-parallel to the wave, when we calculate the particle flux in the inertial frame, we must compare particles of slightly different energy in the wave frame). If we solve this for the diffusion coefficient, we obtain:
\begin{equation}
\kappa(\gamma)=\frac{ f_\tr{p}}{\mathbf{n}\cdot\nabla  f_\tr{p}}\left[-v_\tr{D}-\frac{1}{3}v_\tr{A}\frac{\partial \log  f_\tr{p}}{\partial \log p}\right]\approx L_\tr{z}[v_\tr{D}-\frac{3}{2}v_\tr{A}]
\label{eqn:kappa_stream}
\end{equation}
Here we have set ${\partial \log f_\tr{p}}/{\partial \log p} \approx -4.5$; as before, the energy dependent CR scale length is $L_\tr{z}(\gamma)=| f_\tr{p}/(\mathbf{n}\cdot\nabla  f_\tr{p})|$. If we insert this into the diffusion term in equation \eqref{eqn:crevol}, we obtain:
\begin{equation}
D(r) \equiv \nabla\cdot(\kappa_\tr{p} \mathbf{nn}\cdot\nabla  f_\tr{p}) \approx \nabla \cdot ({f}_\tr{p} \mathbf{n} (v_{\rm D}-v_{\rm A})).
\end{equation}
where we evaluate the drift speed relative to the wave frame, $(v_{\rm D} - v_{\rm A})$, from equations (\ref{eqn:vD_NLLD}) and (\ref{eqn:vD_turb}). Note that the gradient of the distribution function $\nabla  f_\tr{p}$ (or equivalently, the scale height $L_{\rm z}$) does not appear in the diffusion term. It only appears if $(v_{\rm D} - v_{\rm A})$, has a functional dependence on $L_{\rm z}$. This is true for non-linear Landau damping, where $(v_{\rm D} - v_{\rm A}) \propto L_{z}^{-1/2}$ (so that the diffusion term $\propto (\nabla f_\tr{p})^{1/2}$ rather than $(\nabla f_\tr{p})$), but false for turbulent damping, where the diffusion term is therefore {\it independent} of the magnitude of $\nabla f$.

The latter unusual behavior was first noted by \citet{skilling71} for the case of ambipolar damping, which shares similarities with turbulent damping in this regard (although he dismissed it as unimportant, since the effects of diffusion were small for the applications he considered). From equations (\ref{eqn:kappap}) \& (\ref{eqn:nu_collide}), and equating wave growth and with a generic damping rate $\Gamma_{\rm D}$, the diffusion term can be expressed more transparently as \citep{skilling71}:
\begin{eqnarray}
D(r) &=& \frac{1}{p^{3}} \nabla \cdot \left( \frac{\Gamma_\tr{D} B^{2} \mathbf{n}}{4 \pi^{3} m_{p} \Omega_{0} v_{\rm A}} \frac{\mathbf{n}\cdot\nabla f_\tr{p}}{|\mathbf{n}\cdot\nabla f_\tr{p}|}\right) \\
&\approx& \frac{1}{4 \pi^{3} p^{7/2} e^{1/2}m_{p}^{1/2}} \nabla \cdot \left( \frac{B^{3/2} \mathbf{n}}{L_{\rm MHD}^{1/2}}\frac{\mathbf{n}\cdot\nabla f_\tr{p}}{|\mathbf{n}\cdot\nabla f_\tr{p}|} \right)
\label{eqn:diffusion_turb}
\end{eqnarray}
where we specialize to the case of turbulent damping in the second equality, and substitute $\Gamma_{\rm turb} \approx v_{\rm A}/(r_{\rm L} L_{\rm MHD})^{1/2}$. Note that, other than the sign of $\mathbf{n}\cdot\nabla f_\tr{p}$, the term within the divergence is independent of $f_\tr{p}$. This has important consequences for us, in that diffusion does not slow down with time as $\nabla f_\tr{p}$ decreases. Instead, it is independent of $f_\tr{p}$ and depends only on plasma properties. If these plasma properties are roughly constant over the streaming timescale, then $\dot{f_\tr{p}}(r,p,t) \approx D(r,p) \approx$ is roughly constant and $t_{\rm stream} \propto f_\tr{p}/\dot{f_\tr{p}} \propto f_\tr{p}$, with decreases with time as $f_\tr{p}$ falls. This acceleration is key in our more detailed calculations which show that large changes in radio halo luminosity are possible despite apparently long initial diffusion times. It is important to stress, however, that while the diffusion time with turbulent damping is not sensitive to the {\it magnitude} of $\nabla f_\tr{p}$, it is still sensitive to the {\it sign} of $\nabla f_\tr{p}$. The sign of $\mathbf{n}\cdot\nabla f_\tr{p}$ reflects the fact that CRs can only stream along B-fields, down their gradient\footnote{CRs can only stream up a gradient if the sign of energy transfer is reversed -- i.e., the gas gives energy to the CRs, rather than vice-versa, as in Fermi acceleration. In this case, the picture of self-confinement is clearly not applicable.}; diffusion has no further effect if $\nabla f_\tr{p} = 0$. Failure to carefully treat this can result in spurious numerical instabilities \citep{sharma10-stream}, which we discuss in \S\ref{sect:stability}.

\subsection{Collisional Losses}\label{heating}

Cosmic ray protons can also lose energy from direct collisions with gas particles, either through Coulomb interactions, or hadronic interactions (pion production). While these are generally subdominant to losses from wave-particle interactions, we include them for completeness. This transfer of energy from CRs in turn heats the gas.

The energy loss rate of a CR of speed $\beta=v/c$ and kinetic energy $E$ due to Coulomb collisions in ionized gas is: (\cite{mannheim94})
\begin{equation}
\left(\dd{E}{t}\right)_\tr{C}=-4.96\ee{-19}\tr{erg}\ \tr{s}^{-1}\left(\frac{n_\tr{e}}{\tr{cm}^{-3}}\right)\frac{\beta^2}{\beta^3+x_\tr{m}^3}.
\end{equation}
Here $x_\tr{m}=0.0286[T/(2\ee{6}\ \tr{K})]^{1/2}$, with $T$ and $n_\tr{e}$ the gas electron temperature and number density. The energy loss rate of a CR due to hadronic collisions is (\cite{mannheim94}):
\begin{equation}
-\left(\dd{E}{t}\right)_\tr{h}\approx0.5n_\tr{N}\sigma_\tr{pp}\beta cE \, \theta(E-E_\tr{thr})
\end{equation}
where the pp cross section for hadronic interactions is $\sigma_\tr{pp}$ and the target nucleon density is $n_\tr{N}=n_\tr{e}/(1-0.5Y)$, $Y$ being the helium mass fraction. The above assumes an inelasticity of $K=1/2$ for the collision. The Heaviside step function enforces the condition that only cosmic rays with kinetic energy above $E_\tr{thr}=282\ \tr{MeV}$ undergo pion production. All of the energy loss in Coulomb collisions goes toward heating the gas, whereas only $\sim 1/6$ of the inelastic energy in hadronic collisions goes toward secondary electrons which heat the gas, the rest escaping as gamma rays and neutrinos.

These loss terms are represented in the CR Vlasov equation as:
\begin{equation}
\left(\frac{\partial f_\tr{p}}{\partial t}\right)_\tr{C,h}= - \frac{\partial}{\partial p}(\dot{p}_\tr{C,h}f_\tr{p})
\label{eqn:CRcool}
\end{equation}
\[
\dot{p}_\tr{C,h}=\left(\dd{E(p)}{t}\right)_\tr{C,h}\left(\dd{E(p)}{p}\right)^{-1}
\]
where $E=(\sqrt{1+p^2}-1)m_\tr{p}c^2$ and the momentum $p$ is in units of $m_\tr{p} c$.

\subsection{Turbulent Diffusion}
\label{sect:turb}

As we have seen, turbulent gas motions can damp MHD waves and enhance CR streaming. However, they can also directly transport CRs advectively. A proper treatment of the interplay between these effects requires 3D MHD simulations. Here, we will simply treat turbulent motions as a diffusive term in the CR transport equation. If $P_{\rm CR}/P_{\rm gas}$ is small and CRs have negligible effect on the dynamics, they simply act as a passive tracer species. Analogously to the mixing of metals by turbulent diffusion \citep{rebusco06}, we can write:
\begin{equation}
\left( \frac{\partial n_{\rm CR}}{\partial t} \right)_{\rm turb} = - \nabla \cdot \left[ \kappa_{\rm turb} n_{e} \nabla \left(\frac{n_{\rm CR}}{n_{e}} \right) \right],
\label{eqn:CR_diffusion}
\end{equation}
where
\begin{equation}
\kappa_{\rm turb} \approx \frac{v_{t}L_{t}}{3} \approx \frac{v_{A} L_{\rm MHD}}{3}
\label{eqn:kappa_turb}
\end{equation}
i.e., if turbulent mixing is vigorous, the CRs will have uniform relative abundance, $n_{\rm CR} \propto n_{e}$. This has some support from simulations where CR dynamics are taken into account \citep{sharma09}. There, turbulent convection results in constant CR entropy $P/n_{\rm CR}^{\gamma_{\rm CR}}$ (where $\gamma_{\rm CR} = 4/3$) and $P_{\rm CR}/P_{\rm g}=$const. This implies $n_{\rm CR} \propto P_{\rm g}^{1/\gamma_{\rm CR}}$. Since stratified gas in a cluster has a polytropic equation of state $P_{\rm g} \propto \rho_{g}^{\gamma_{\rm pt}}$ where $\gamma_{\rm pt} \approx 1.2-1.3$ (e.g., \citet{capelo12}), this implies $n_{\rm CR} \propto (\rho_{g})^{\gamma_{\rm pt}/\gamma_{\rm CR}} \propto \rho_{g}^{0.9-0.98}$, consistent with our assumptions. Alternatively, \citet{enslin11} suggest a target profile set by gas entropy, rather than CR entropy: $n_{\rm CR} \propto P_{g}^{1/\gamma_{g}}$, where $\gamma_{g} = 5/3$. In this case, all occurrences of $n_{e}(r)$ in equation (\ref{eqn:CR_diffusion}) will be replaced by $\eta(r)=P_{g}^{1/\gamma_{g}}$, and $n_{\rm CR} \propto \rho_{g}^{\gamma_{\rm pt}/\gamma_{g}} \propto \rho_{g}^{0.72-0.78}$. Given the many uncertainties in the model, this difference in scalings is of secondary importance.

We also need to take adiabatic heating and cooling into account. The normalization of the distribution function $f$ varies with adiabatic changes as $C \propto n_{\rm CR}^{\alpha/3}$ (e.g., \citet{enslin07}), where $\alpha=4-5$ is the spectral slope of the distribution function. Thus, the overall effect of turbulent diffusion on the distribution function is:
\begin{equation}
\frac{\partial f}{\partial t} = - \nabla \cdot \left[ \kappa_{\rm turb} \delta^{\alpha/3} \nabla \left(\frac{f}{\delta^{\alpha/3}} \right) \right],
\label{eqn:f_diffusion}
\end{equation}
and $\delta (r) = P_{\rm g}^{1/\gamma_{\rm CR}} \approx n_{e}(r)$, or $\delta(r)= \eta(r) = P_{g}^{1/\gamma_{g}}$.

These equations show that turbulent diffusion, acting alone, will lead to a centrally peaked CR profile similar to the gas profile. On the other hand, turbulence also damps MHD waves, leading to enhanced outward streaming, which flattens the CR profile. Which effect dominates? While we explore this in detail in our numerical calculations, it is useful to first get an order of magnitude estimate. From equations (\ref{eqn:kappa_stream}) and (\ref{eqn:kappa_turb}), we obtain:
\begin{equation}
\frac{\kappa_{\rm stream}}{\kappa_{\rm turb}} \approx \left( \frac{v_{\rm D}}{v_{\rm A}} - 1 \right) \left( \frac{L_{\rm z}}{L_{\rm MHD}} \right) \propto \frac{1}{L_{\rm MHD}^{3/2}}
\end{equation}
We expect $L_{\rm z}/L_{\rm MHD} \gsim 1$, and $(v_{\rm D}/v{\rm A} -1) \gsim 1$ (from equation \eqref{eqn:vD_turb}) for turbulent damping; moreover, these factors increase during the streaming processes as $L_{\rm z}$ rises and $n_{\rm CR}$ fall. Thus, $\kappa_{\rm stream} \gsim \kappa_{\rm turb}$ in our fiducial model. Moreover, if the strength of turbulence increases such that $L_{\rm MHD}$ falls, $\kappa_{\rm stream}/\kappa_{\rm turb}$ {\it rises}. Stronger turbulence has a larger effect on damping of MHD waves than on inward advection of CRs, and the CRs stream outward faster. Thus, in this framework, turbulent diffusion can never establish a centrally peaked profile, regardless of its strength. In practice, coherent bulk motions (triggered by mergers, or perhaps by gas `sloshing') can potentially bring CRs to the cluster center, and/or produce a magnetic topology which is unfavorable for outward streaming. However, modeling such stochastic events is beyond the scope of this paper.

A few comments about our choice of fiducial parameters for turbulence is in order. It is customary to define $(L_{\rm t},v_{\rm t})$, where $L_{\rm t}$ is the outer scale, and $v_{\rm t}$ is the velocity at this scale. Instead, we work with $(L_{\rm MHD}, v_{\rm A})$, where $L_{\rm MHD}$ is {\it defined} to be the scale at which the turbulent velocity is $v_{\rm A}$. In general, $v_{t} \sim v_{\rm A} (L_{\rm t}/L_{\rm MHD})^{1/3}$, and more vigorous turbulence can be characterized by smaller values of $L_{\rm MHD}$. However, if there is equipartition between $U_{\rm B} = B^{2}/8\pi$ and $U_{t} = 1/2 \rho v_{t}^{2}$, then $v_{\rm t} \sim v_{\rm A}$ and thus $L_{\rm t} \sim L_{\rm MHD}$. Thus, $(L_{\rm MHD}, v_{\rm A})$ are sensible fiducial parameters. Secondly, we have assumed that $L_{\rm MHD}$ (or equivalently, $v_{t}$ at a fixed scale) is independent of radius. Is this consistent with cosmological simulations, which show that turbulent pressure support becomes increasingly important with radius? A fit to low-redshift clusters gives (\citet{shaw10}, see also \cite{battaglia12}):
\begin{equation}
\left( \frac{P_{\rm turb}}{P_{\rm therm}} \right) = \alpha_{0} \left( \frac{r}{R_{500}} \right)^{n_{\rm nt}}
\end{equation}
where $\alpha_{0} \approx 0.18 \pm 0.06$ and $n_{\rm nt} = 0.8 \pm 0.25$. This implies $v_{\rm t} \propto r^{0.4-\alpha_{T}/2}$, where $T\propto r^{-\alpha_{T}}$, and $\alpha_{\rm T}$ is generally small (e.g., $\alpha_{T} \approx x^{2}/(1+1.5x)$ \citep{loken02}, where $x\equiv r/r_{\rm vir}$, so $\alpha_{\rm T} \approx 0.2$ at $r=0.5 r_{\rm vir}$). On the other hand, given our assumption that $B \propto \rho^{\alpha_{\rm B}}$ (see \S\ref{sect:initial_conditions}), we have $v_{\rm A} \propto r^{(0.5-\alpha_{\rm B})\alpha_{\rho}}$, where $\rho \propto r^{-\alpha_{\rho}}$, and $\alpha_{\rho} \approx 2-3$ over most of the cluster. The radial scalings for $v_{\rm t}$ and $v_{\rm A}$ are thus roughly consistent: for instance, $\alpha_{\rm B} \approx 0.3$ (as assumed for Perseus \& Coma) gives $v_{\rm A} \propto r^{0.4-0.6}$. Finally, we note that for our assumed levels of turbulence, heat dissipation is relatively unimportant. The heating time is:
\begin{equation}
t_{\rm heat} \sim \frac{U_{\rm therm}}{\epsilon} \sim t_{\rm turb} \left(\frac{U_{\rm therm}}{U_{\rm turb}} \right)  \sim 5 \, {\rm Gyr} \frac{f_{\rm t, 5} L_{\rm MHD,100}}{v_{\rm A,100}} %\left( \frac{U_{\rm therm}/{U_{\rm turb}}{5} \right)
\end{equation}
where $f_{t,5}=[(U_{\rm therm}/U_{\rm turb})/5]$, $L_{\rm MHD,100}=(L_{\rm MHD}/100$ kpc), and $v_{\rm A,100} = (v_{A}/100 \, {\rm km \, s^{-1}})$.

\section{Method}
\label{sect:method}

Our main task is to solve the CR transport equation, in the form:
\begin{equation}\label{crevol}
\begin{split}
\frac{\partial  f_\tr{p}}{\partial t}+(\mathbf{u}+\mathbf{v}_\tr{A})\cdot\nabla  f_\tr{p}=\nabla\cdot(\kappa_\tr{p} \mathbf{nn}\cdot\nabla  f_\tr{p})\\ +\frac{1}{3}p\frac{\partial  f_\tr{p}}{\partial p} \nabla\cdot (\mathbf{u}+\mathbf{v}_\tr{A})+ {Q} - \frac{\partial}{\partial p}(\dot{p}_\tr{C,h}f_\tr{p}) \\ - \nabla \cdot \left[ \kappa_{\rm turb} \delta^{\alpha/3} \nabla \left(\frac{f}{\delta^{\alpha/3}} \right) \right]
\end{split}
\end{equation}
where the last two terms are as in equations (\ref{eqn:CRcool}) and (\ref{eqn:f_diffusion}) respectively. To do so, we have written a new module in a 1D spherically symmetric version of ZEUS3D, previously used to solve the CR equations in the fluid approximation \citep{Guo08a}.

Our goal in this paper is to determine if CR streaming is a plausible means of turning off radio halos in the hadronic scenario. We therefore run numerical simulations where the cluster is assumed to be in strict hydrostatic and thermal equilibrium, and only solve the CR transport equation (ignoring the fluid equations for the gas, equations , by setting all time derivatives to zero) to examine the effects of CR streaming. In the absence of a cooling flow, the only time-dependent terms in the fluid equations for the gas (equations (\ref{masscon}),(\ref{momcon}), and (\ref{enercon})) relate to the CRs, and have negligible effect. We initialize the CR profile so as to reproduce the observed radio surface brightness profiles in the classical hadronic model, and follow the time evolution of radio emission as the CRs stream out. In this methods section, we discuss numerical regularization of CR streaming (\S\ref{sect:stability}), a test comparison of our CR transport solver in the fluid approximation (\S\ref{sect:test_AGN}), calculating radio and gamma-ray emission (\S\ref{sect:emissivity}), and our initial conditions (\S\ref{sect:initial_conditions}) for a prototypical radio mini-halo (Perseus) and a prototypical giant radio halo (Coma). Results are then presented in the following section, \S\ref{results}.

\subsection{CR streaming: Numerical Stability}
\label{sect:stability}

Cosmic rays can only stream down their gradient, in a direction:
\begin{equation}
\mathbf{s}=-{\rm sgn}(\mathbf{B} \cdot \nabla f_\tr{p}) \frac{\mathbf{B}}{|{\mathbf B}|}.
\end{equation}
In our 1D simulations, $\mathbf{s}= -{\hat{\mathbf r}}\, {\rm sgn}({\rm d}f_\tr{p}/{\rm d}r)$. However, if this is enforced in equation (\ref{eqn:diffusion_turb}), it leads to numerical instabilities and unphysical oscillations in the distribution function. The origin of this difficulty is easy to understand \citep{sharma10-stream}; it essentially arises at local extrema. If the simulation at any time produces a local density maximum, diffusion out of the local maximum will cause the density to drop significantly there. If the time step is not properly restricted, this decrease will overshoot, causing the density to drop below neighboring regions, creating a local minimum. The opposite problem will then occur, with inwardly diffusing CRs causing the CR density to increase too much. The result is an unphysical oscillation that eventually spreads out to all space. Because CR streaming results in a flat profile where $({\rm d}f/{\rm d}r)$ vanishes everywhere, this problem can become acute as time goes on.

\citet{sharma10-stream} show that for an explicit code (such as ZEUS3D), the restriction on the time-step such that new local extrema are not created is:
\begin{equation}\label{time0}
  \Delta t\le|f''|\Delta x^3/f|v|
\end{equation}
which is much more onerous than the standard Courant condition (we have explicitly verified that simulations which satisfy the Courant condition suffer from spurious oscillations). They suggest regularizing the CR transport equation by replacing {the discontinuous} $\tr{sgn}(f'_\tr{p})$ with {the smooth function} $\tr{tanh}(f'_\tr{p}/\epsilon)$ {for some choice of $\epsilon$}. As $\epsilon$ tends to zero, the tanh function approaches the sign function. This effectively sets $\epsilon$ as a minimum scale value for $f'_\tr{p}$: if $f'_\tr{p} \ll \epsilon$, the simulation behaves as if $f'_\tr{p}=0$, and suppresses CR streaming. Alternatively, it can be viewed as introducing a diffusive term at an extremum, with diffusion coefficient $f_\tr{p}/\epsilon$, similar to the use of explicit viscosity to regularize Euler/Burger's equations. In this case, the maximum time step allowed to suppress the instability is:
\begin{equation}\label{time1}
 \Delta t\le\Delta x^2\epsilon/2f_p|v|.
\end{equation}
For us, the relevant speed $v$ is the streaming speed, calculated via equation (\ref{vDrelation}), which we insert into this equation.

\begin{figure}
\includegraphics[width=8.5cm,trim=2cm 10cm 4cm 4cm]{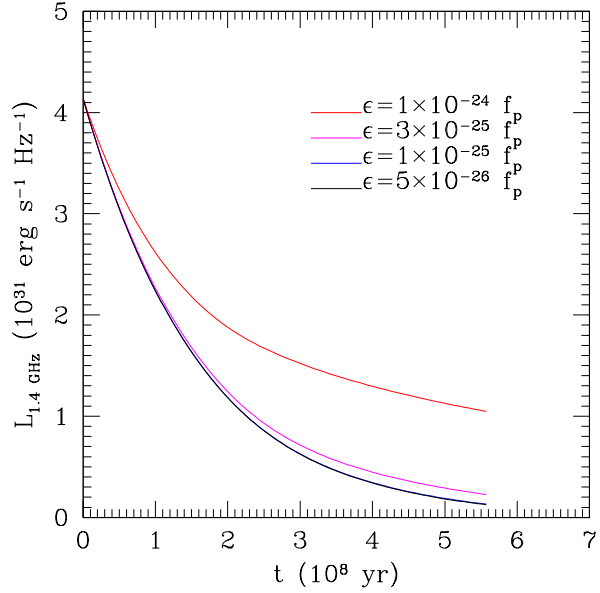}
\caption{Convergence test for the smoothing scale parameter $\epsilon$. We plot the 1.4 GHz radio luminosity of Perseus, which declines with time due to cosmic-ray streaming (here we assume $L_{\rm MHD} = 400$kpc; see \S\ref{Perseus} for details). As $\epsilon$ is decreased, our calculations converge. Our fiducial value is $\epsilon = 10^{-25} f_{\rm p}$.}
\label{eps}
\end{figure}

We have found that a scale value of:
\begin{equation}
 \epsilon=f_p/L \, ; \ \ L = 3 \, {\rm Mpc}
\end{equation}
is sufficiently small that decreasing it any further does not significantly change the results. In Fig \ref{eps}, we show a convergence test (showing the radio luminosity of Perseus as a function of time when \textbf{$L_{\rm MHD}=100$ kpc}; see Fig \ref{perseusplots}b) where the figure converges to the correct solution as $\epsilon$ is decreased; a value of $\epsilon$ half of our fiducial value ($\epsilon = 10^{-25} f_{\rm p}$) gives identical results.

{In practice, although we use a smoothing scale $\epsilon$, we use the time constraint \eqref{time0} rather than \eqref{time1} and we use $v=v_\tr{A}$ rather than $v=v_\tr{s}$. Additionally, to prevent the time step from dropping to zero we impose a minimum time step
\begin{equation}\label{time2}
\Delta t\ge1\ee{-7}\Delta x^2\epsilon/2f_p|v|
\end{equation}
where the numerical factor out front is arbitrary. We do all of this to regulate the runtime of the simulation - the less stringent time steps will be less accurate but will run quicker, and we can adjust the minimum time step \eqref{time2} to the desired balance of speed and accuracy. As a result, local minima and maxima do develop at some points of our simulations, however this will always happen as the CR profile flattens; as long as the local extrema do not grow unstably the results should be robust.}

In addition, we enforce the constraint that $\Delta f_\tr{p} \leq 0.05 f_\tr{p}$ in a single time-step {(and similarly for the gas energy and density). Note that while \eqref{time0} is only applied at local extrema, this condition is held everywhere.} Eventually, as $f_\tr{p}$ falls, this shrinks the time-step to zero. To avoid this, we define a momentum dependent minimum ${f}_\tr{p,min}(p) = 10^{-3} f_\tr{p}(r_{\rm max},p, t_{\rm 0})$, where $r_{\rm max}$ is the outer boundary of the simulation, and $t_{\rm 0}$ is the initial time. {The distribution function $f_\tr{p}$ is then never allowed to drop below this value}. Also, once $f_\tr{p}$ falls below $50 {f}_\tr{p,min}(p)$ {anywhere, all time step restrictions there are ignored, including \eqref{time0}.}

\subsection{Test Case: AGN Feedback}
\label{sect:test_AGN}

To test our solver for the CR transport equation, it is useful to compare against previous results where CRs are treated in the fluid approximation. Specifically, we compare against the results of \citet{Guo08a}, which simulates the effects of CRs injected by a central AGN on the thermal state of a cool core cluster. It was found that a combination of electron thermal conduction (at some fraction $f$ of the Spitzer value) and CR mediated wave heating was sufficient to stem a cooling flow. The following governing equations for the two-fluid ICM (gas and cosmic rays) were used:
\begin{equation}\label{masscon}
\frac{\partial\rho}{\partial t}+\nabla\cdot(\rho \mathbf{u})=0
\end{equation}
\begin{equation}\label{momcon}
\frac{\partial \mathbf{S}}{\partial t}+\nabla\cdot(\mathbf{S}u)=-\nabla P_\tr{g}-\nabla P_\tr{c}-\rho\nabla\Phi
\end{equation}
\begin{equation}\label{enercon}
\begin{split}
\frac{\partial E_\tr{g}}{\partial t}+\nabla\cdot(E_\tr{g}\mathbf{u})&=-P_\tr{g}\nabla\cdot\mathbf{u}-\nabla\cdot \mathbf{F}\\
&-n_\tr{e}^2\Lambda(T)+\eta_\tr{c}n_\tr{e}E_\tr{c} -\mathbf{v}_\tr{A}\cdot\nabla P_\tr{c}
\end{split}
\end{equation}
\begin{align}
\frac{\partial E_{\text{c}}}{\partial t} &=(\gamma_{\text{c}}-1)
(\mathbf{u}+\mathbf{v}_{\text{A}})\mathbf{\cdot \nabla} E_{\text{c}}
- \mathbf{\nabla \cdot F}_{\text{c}} + {Q}_{c}    \text{.} \label{eqn:CRold1} \\
\mathbf{F}_{\text{c}}&=\gamma_{\text{c}}E_{\text{c}}(\mathbf{u}+\mathbf{v}_{\text{A}})- \mathbf{n}\kappa_{\text{c}}(\mathbf{n \cdot \nabla} E_{\text{c}}) \text{,} \label{eqn:CRold2}
\end{align}
where $\rho $ is the gas density, $P_{\text{g}}$ is the gas pressure, $E_{\text{g}}$ is the gas energy density, $ \mathbf{S}= \rho \mathbf{u}$ is the gas momentum vector, $E_{c}$ is the cosmic ray energy, $P_{\text{c}}=(\gamma_{\text{c}}-1)E_{\text{c}}$ is the cosmic-ray pressure, ${\mathbf F}$ is the electron conduction heat flux, and $\Phi$ is the gravitational potential. The term $\eta_\tr{c}n_\tr{e}E_\tr{c}$, where $\eta_{\text{c}} =2.63 \times 10^{-16}\text{ cm}^{3}\text{ s}^{-1}$, takes Coulomb and hadronic heating of the gas by cosmic rays into account. The initial conditions, gravitational potential $\Phi(r)$, and cooling function $\Lambda(T)$ are as spelled out in \citet{Guo08a}; please refer to the paper for details. The source function ${Q}$ represents the injection of CRs by an AGN, triggered by gas cooling:
\begin{equation}\label{source}
Q_\tr{c}=-\frac{\nu\epsilon\dot{M}_\tr{in}c^2}{4\pi r_0^3}\left(\frac{r}{r_0}\right)^{-3-\nu}[1-e^{-(r/r_0)^2}]
\end{equation}
Here, $\epsilon=3 \times 10^{-3}$ is an efficiency parameter, $\nu=0.3$, and $r_0=20\ \tr{kpc}$ is a scale distance.

The code of \citet{Guo08a} uses the CR energy density $E_{c}$ as the fundamental dynamic variable for CRs. It is:
\begin{equation}
E_\tr{c}=4\pi\int_0^\infty p^2T_\tr{p}(p_\tr{p}) {f}_\tr{p}(p_\tr{p}) \tr{d}p
\label{eqn:Ec}
\end{equation}
where
\begin{equation}
T_{\text{p}}(p)=\biggl[\sqrt{1+p^{2}}-1\biggr]mc^{2}  \text{.}
\label{eqn:Tc}
\end{equation}
is the kinetic energy of a CR proton of momentum $p_\tr{p}$. By using $E_{c}$ as the main CR dynamic variable, all momentum dependence has been integrated out. By contrast, we wish to retain momentum dependence, and instead use $f(r,p,t)$ as our fundamental variable. We therefore continue to solve equations $\eqref{masscon}-\eqref{enercon}$, but replace equations (\ref{eqn:CRold1}) \& (\ref{eqn:CRold2}) with the equation for the distribution function, equation (\ref{crevol}), and solve for $E_{c}$ as required in equations (\ref{momcon}), (\ref{enercon}) via equations (\ref{eqn:Ec}) and (\ref{eqn:Tc}). In calculating the momentum-dependent source function ${Q}$ for use in equation (\ref{crevol}), we suppose that the injected spectra has the form:
\begin{equation}\label{initialf}
f_\tr{p}(E)=\frac{A_\tr{cr}\theta(E-E_\tr{l})}{(E/E_*)^{\tilde{\alpha}}+(E/E_*)^{(\tilde{\alpha}-2)/2}}
\end{equation}
which assumes a steady state spectrum at low (high) energies due to Coulomb (hadronic) losses, and smoothly connects these regimes (see \citep{Guo08a} for details).  Here, $E_*=706\ \tr{MeV}$ is a cross-over energy separating the low- and high-energy regimes, while the assumed spectral index is $\tilde{\alpha}=2.5$, and cutoff energy is $E_\tr{l}=10\ \tr{MeV}$. We create a source function with the same momentum dependence as equation (\ref{initialf}), and then normalize it to the total CR injection rate given by equation (\ref{source}):
\begin{equation}
Q_\tr{c}=4\pi\int_0^\infty p^2T_\tr{p} Q \, \tr{d}p.
\end{equation}
To maintain consistency with \citet{Guo08a}, we use the same momentum-independent diffusion coefficient used there.

The simulation grid has two ghost zones at each end in the radial direction, and one ghost zone at each end in the momentum direction. The density and temperature of the ICM are linearly extrapolated into the spatial ghost zones. For the CR spectrum, constant boundary conditions are enforced in the radial direction, i.e. ${f_\tr{p}}(p)$ at the spatial ghost zones are set equal to ${f_\tr{p}}(p)$ in the adjacent active zones. In the momentum direction we required that $\tr{d}\log f_\tr{p}/ \tr{d}\log p$ be constant across the boundary. As for time step constraints, for this test case we do not allow $\rho_{\rm g}$ or $E_{g}$ to change by more than $25\%$ in any time step. We also enforce the Courant condition for all cells, $\Delta t < \Delta x^{2}/2 \kappa_{\rm p}$.

\begin{figure}
\includegraphics[width=8.5cm,trim=2cm 10cm 4cm 4cm]{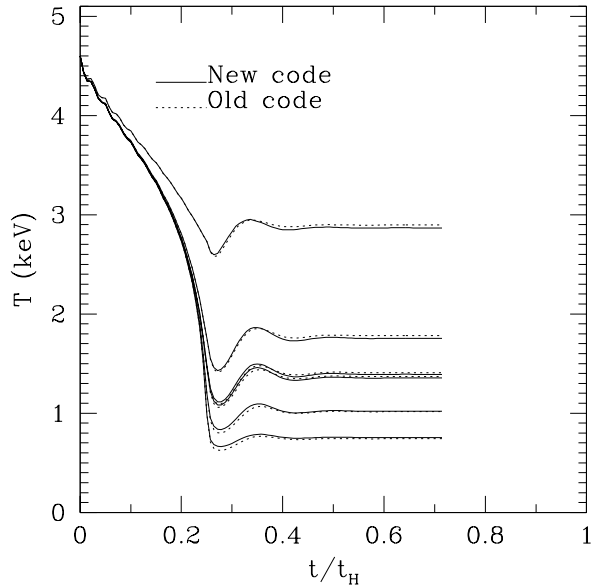}
\caption{Temperature versus simulation time at some select radii for the new code, where we solve the CR transport equation (\ref{crevol}). The results of the old code, which treats CRs in the fluid limit (equations \eqref{eqn:CRold1}, \eqref{eqn:CRold2}), are displayed in the dotted lines.}
\label{fig:comparison}
\end{figure}

We find that our full model reproduces the results of \citet{Guo08a} extremely well. A example is shown in Fig. \ref{fig:comparison}, where we show the temperature as a function of time for several select radii. The cluster is initialized to be isothermal; after an initial transient, it is thermally stabilized against a cooling catastrophe by a combination of CR heating and electron thermal conduction. The dotted lines show the results from the code of \citet{Guo08a}, which integrates the fluid equations, while the solid lines indicate the results of the new code, which computes the distribution function. Indeed, even when we include the full momentum dependence of the diffusion coefficient, the results barely change (for this particular example, we have assumed non-linear Landau damping). This is because most of the energy density of CRs is dominated by low energy CRs ($\sim 1 \, {\rm GeV}$) for which the diffusion time is negligibly long. As we shall soon see, diffusion cannot be neglected for the high-energy CRs which are responsible for observed radio emission.

\subsection{Computing Emissivities}
\label{sect:emissivity}

Our fundamental simulation variable is the CRp distribution function, $f_{\rm p}(r,p,t)$. Here, we describe how radio and gamma-ray emission can be inferred from $f_{\rm p}(p)$ (hereafter, we suppress $r,t$) in the hadronic model, given an assumed gas density and magnetic field. %Because the momentum dependence of the diffusion coefficient introduces curvature in $f_{\rm p}(p)$, we do not assume that it is necessarily a power-law, but use more general expressions.

How is radio emission produced? CR protons undergo hadronic interactions to produce pions, which in turn decay to produce relativistic electrons (CRp + nucleon $\rightarrow \pi^{\pm},\pi^{0}; \ \pi^{\pm} \rightarrow \mu^{\pm} + \nu_{\mu}/\bar{\nu}_{\mu} \rightarrow e^{\pm} + \nu_{e}/\bar{\nu}_{e} + \nu_{\mu}+\bar{\nu}_{\mu}; \ \pi^{0} \rightarrow 2 \gamma$). The high energy electrons which produce observable synchrotron emission have short cooling lifetimes and we therefore assume a steady-state between injection and cooling. A CRp distribution $f_\tr{p}(p)$ gives rise to a pion source function due to the hadronic pp interaction \citep{pfrommer08a}:
\begin{equation}
\begin{split}
s_{\pi^\pm}(p_\pi)&=\frac{2}{3}\int_{-\infty}^\infty\tr{d}p_\tr{p}f_\tr{p}(p_\tr{p})cn_\tr{N}\xi(p_\tr{p})\sigma_\tr{pp}^\pi\\
&\times\delta\left(p_\pi-\frac{m_\tr{p}}{4m_\pi}p_\tr{p}\right)\theta(p_\tr{p}-0.78)
\end{split}
\end{equation}
where $\sigma_\tr{pp}^\pi=32(0.96+e^{4.4-2.4\alpha_\tr{p}})$ mbarn \footnote{We follow \citet{pfrommer04} in absorbing the weak energy dependencies of the
pion multiplicity and the inelastic cross-section in this semi-analytical parametrization of the cross-section, where $\alpha_\tr{p}$ is the average CR spectral index.}. The delta function enforces the mean pion momentum $\langle \tilde{p}_\pi\rangle=\tilde{p}_\tr{p}/4$ and the Heaviside step function $\theta$ incorporates the threshold proton momentum for the pion production to occur. Approximating the pion multiplicity $\xi$ as 2, this gives:
\begin{equation}
\begin{split}
s_{\pi^\pm}(p_\pi)=&\frac{16}{3}\frac{m_\pi}{m_\tr{p}}cn_\tr{N}\sigma_\tr{pp}^\pi 4 \pi \left(\frac{4m_\pi}{m_\tr{p}}p_\pi\right)^{2}\\
&\times f_\tr{p}\left(\frac{4m_\pi}{m_\tr{p}}p_\pi\right)\theta\left(\frac{4m_\pi}{m_\tr{p}}p_\pi-0.78\right)
\label{eqn:pion_source}
\end{split}
\end{equation}
Under this approximation, the neutral pion source function $s_{\pi^0}$ is the same.

The charged pion population will undergo pion decay, producing electrons (and other particles). This electron production is described with the electron source function:
\begin{equation}
s_\tr{e}(p_\tr{e})=s_{\pi^\pm}(p_\pi(p_\tr{e}))\dd{p_\pi}{p_\tr{e}}
\end{equation}
\begin{equation}\label{fe1}
\begin{split}
\Rightarrow s_\tr{e}(p_\tr{e})=&\frac{64}{3}\frac{m_\tr{e}}{m_\tr{p}}cn_\tr{N}\sigma_\tr{pp}^\pi 4 \pi \left(\frac{16m_\tr{e}}{m_\tr{p}}p_\tr{e}\right)^{2} \\
&\times f_\tr{p}\left(\frac{16m_\tr{e}}{m_\tr{p}}p_\pi\right)\theta\left(\frac{16m_\tr{e}}{m_\tr{p}}p_\tr{e}-0.78\right)
\end{split}
\end{equation}
In the second equation we have used $\tilde{p}_\pi= 4 \tilde{p}_{e}$. If we assume an equilibrium between this source and any losses, i.e. a steady state solution, the electron spectrum is then determined from
\begin{equation}\label{fe2}
f_\tr{e}(p_\tr{e})=\frac{1}{|\dot{p}_\tr{e}|}\int_{p_\tr{e}}^\infty \tr{d}p_\tr{e}'s_\tr{e}(p_\tr{e}')
\end{equation}
where the losses $\dot{p}_\tr{e}$ are
\begin{equation}
\dot{p}_\tr{e}(p_\tr{e})=\frac{\dot{E}_\tr{e}}{m_\tr{e}c^2}=\frac{4}{3}\frac{\sigma_\tr{T}cp_\tr{e}^2}{m_\tr{e}c^2}(\varepsilon_\tr{B}+\varepsilon_\tr{cmb})
\end{equation}
from synchrotron radiation and inverse Compton (IC) scattering. Here, $\sigma_{\rm T}$ is the Thompson scattering cross section, and $\varepsilon_{\rm B}, \varepsilon_{\rm cmb}$ are the energy density of the B-field and cosmic microwave background.

From the electron distribution function we can determine the resulting synchrotron emissivity \citep{rybicki79}:
\begin{equation}\label{empower}
j_\nu(r)=0.333\frac{\sqrt{3}}{2\pi}\frac{e^3 B(r)}{m_\tr{e}c^2}\int_1^\infty\tr{d}\gamma_\tr{e} f_\tr{e}(r,\gamma_\tr{e})F\left(\frac{\nu}{\nu_\tr{c}}\right)
\end{equation}
In the above, $\nu_\tr{c}=3eB\gamma_\tr{e}^2/4 \pi m_\tr{e}c$ and the function $F$ is an integral of a modified Bessel function, $F(x)=x\int_x^\infty K_{5/3}(x')\tr{d}x'$. The numerical factor in front comes from averaging the CRe population over pitch angle, assuming isotropy. The observed surface brightness is:
\begin{equation}\label{surface}
S_{\nu} (r_\perp)=\int_{-\infty}^{\infty}j_{\nu^{\prime}}(r(l))\tr{d}l=\frac{2}{(1+z)^{3}}\int_{r_\perp}^{\infty}j_{\nu^{\prime}}(r)\frac{r \, \tr{d}r}{\sqrt{r^2-r_\perp^2}}
\end{equation}
where $\nu^{\prime}=\nu(1+z)$. The luminosity is:
\begin{equation}
%\begin{split}
L_{\nu}= \int \tr{d}^3\mathbf{r}j_\nu(r)
%\end{split}
\end{equation}
%Above, the angular diameter distance $d_a$ is related to the luminosity distance $D$ by $D=(1+z)^2d_a$.

%Note that if we insert a power law for $f_\tr{e}$ into \eqref{empower} above, we recover equation \eqref{jeRelation}. Similarly, inserting a power law for $f_\tr{p}$ into \eqref{fe1} and \eqref{fe2} will reproduce equation \eqref{peRelation}.

We also make predictions for gamma-ray emission. We only consider gamma-ray emission from neutral pion decay $\pi^0\rightarrow2\gamma$ and ignore the subdominant contribution from inverse Compton scattering. The analysis is much the same as above. Following \cite{mannheim94}, we derive a photon source function from the pions:
\begin{equation}
s_\gamma(E_\gamma)=2\int^\infty_{E_\gamma+\frac{(m_\pi c^2)^2}{E_\gamma}}\frac{\tr{d}E_\pi s_{\pi^0}(E_\pi)}{\sqrt{E_\pi^2-m_\pi^2c^4}}
\end{equation}
where the neutral pion source function $s_{\pi^0}(E_\pi)$ is assumed to be the same as for charged pions, equation (\ref{eqn:pion_source}). From this source function, we determine a number production rate per unit volume $\lambda_\gamma$:
\begin{equation}
\lambda_\gamma(>E_\gamma)=\int_{E_\gamma'}^\infty \tr{d}E_\gamma' s_\gamma(E_\gamma')
\label{gammaemission}
\end{equation}
and the flux detected at Earth above an energy $E_\gamma$:
\begin{equation}
F_\gamma(>E_\gamma)=\frac{1}{4\pi d_{\rm L}^2}\int\tr{d}^3\mathbf{r}\lambda_\gamma(>E_\gamma)
\label{eqn:gammaflux}
\end{equation}

Given the strong momentum dependence of CR streaming, it is worth clarifying which range of CRp momenta are most observationally relevant. For radio emission, the characteristic synchrotron frequency is $\sim 3 \gamma^{2} \nu_{c}$, where $\nu_{c}$ is the non-relativistic synchrotron frequency. For a given observational frequency $\nu_{s}$, the greatest contribution comes from electrons with:
\begin{equation}
p_{\rm emit} \approx \gamma_{\rm emit} \approx 4 \times 10^{3} \left( \frac{\nu_{s}}{1 \, {\rm GHz}}\right)^{1/2} \left( \frac{B}{3 \, \mu{\rm G}} \right)^{-1/2}
\end{equation}
Thus, $\sim 10$ GeV electrons are responsible for $\sim$GHz emission in $\mu$G fields. Typically, $\tilde{p}_{e} \sim (1/16) \tilde{p}_{\rm p}$, where $\tilde{p}_{e}$ is the momentum of a secondary CRe produced hadronically. The reduction in energy by a factor of $\sim 16$ comes from the fact that the limiting inelasticity is $\sim 1/2$ \citep{mannheim94}, the pion multiplicity is a factor of $\sim 2$ due to 2 pion jets leaving the interaction site \citep{nachtmann90}, and $\langle E \rangle = (1/4) \langle E_{\pi^{\pm}} \rangle$ in the reaction $\pi^{\pm} \rightarrow e^{\pm} + 3 \nu$. Thus for $\sim {\rm GHz}$ emission, $\sim 100$ GeV protons are most relevant, while for LOFAR observations at $\sim 100$ MHz, $\sim 10$ GeV protons are most relevant. For $\gamma$-ray emission, $E_{\gamma} \approx (1/8) E_{p}$ (all the factors are as before, except $E_{\gamma} = 1/2 E_{\pi^{0}}$). Thus, Fermi, which is most sensitive in the $E_{\gamma} \approx 0.1-3$ GeV range (rather than $0.1-300$ GeV, due to the pion bump), probes $E_{\rm p} \sim 1-30$ GeV, while imaging air Cerenkov telescopes such as MAGIC, HESS and VERITAS are most sensitive in the $E_{\gamma} \sim 0.3-1$ TeV range (rather than 0.3-10 TeV, due to the steep CRp spectrum), probes $E_{\rm p} \sim 3-10$ TeV.

\subsection{Initial and Boundary Conditions}
\label{sect:initial_conditions}

We choose to simulate a prototypical radio mini-halo, Perseus, and a prototypical giant radio halo, Coma. We choose initial conditions which reproduce current observations of their radio surface brightness, and then watch how this evolves under the influence of streaming. For the Perseus cluster, we adopt empirical fits to the cluster temperature and electron density profiles (\cite{pinzke10}) based on observed X-ray emission (\cite{churazov03}):
\begin{equation}
\begin{split}
\frac{n_\tr{e}}{10^{-3}\tr{cm}^{-3}}&=46\left[1+\left(\frac{r}{57\ \tr{kpc}}\right)^2\right]^{-1.8}\\
&\qquad+4.79\left[1+\left(\frac{r}{200\ \tr{kpc}}\right)^2\right]^{-0.87}
\end{split}
\end{equation}
\begin{equation}
T=7\ \tr{keV}\frac{1+(r/71\ \tr{kpc})^3}{2.3+(r/71\ \tr{kpc})^3}\left[1+\left(\frac{r}{380\ \tr{kpc}}\right)^2\right]^{-0.32}
\end{equation}
From these we determine an internal energy distribution (via the ideal gas law) and a gravitational potential (via hydrostatic equilibrium). Similarly, for Coma the fits are (\citet{pinzke10}, based on \citet{briel92}):
\begin{equation}
\frac{n_\tr{e}}{10^{-3}\tr{cm}^{-3}}=3.4\left[1+\left(\frac{r}{294\ \tr{kpc}}\right)^2\right]^{-1.125}
\end{equation}
\begin{equation}
T=8.25\ \tr{keV}\left[1+\left(\frac{r}{460\ \tr{kpc}}\right)^2\right]^{-0.32}
\end{equation}
The radio surface brightness profiles at 1.4 GHz are fit by a $\beta$ profile:
\begin{equation}
S(r)=S_0[1+(r/r_\tr{c})^2]^{-3\beta+0.5}
\end{equation}\label{sbfit}
which is reproduced by an initial emissivity of
\begin{equation}\label{radiofit}
j_\nu(r)=j_{\nu,0}[1+(r/r_\tr{c})^2]^{-3\beta}
\end{equation}
\begin{displaymath}
j_{\nu,0}=\frac{S_0}{2\pi r_c}(6\beta-1)\mathcal{B}\left(\frac{1}{2},3\beta\right)
\end{displaymath}
where $\mathcal{B}$ is the beta function. For Perseus, $\beta=0.55$, $r_c=30\ \tr{kpc}$, and $S_0=2.3\ee{-1}\ \tr{Jy}\ \tr{arcmin}^{-2}$ \citep{pedlar90}, while for Coma $\beta=0.78$, $r_\tr{c}=450\ \tr{kpc}$, and $S_0=1.1\ee{-3}\ \tr{Jy}\ \tr{arcmin}^{-2}$ \citep{deiss97}.

We assume that the magnetic field scales with gas density:
\begin{equation}
B=B_0\left(\frac{n_\tr{e}(r)}{n_e(0)}\right)^{\alpha_B}
\end{equation}
Such a scaling is motivated by simulations \citep{dubois08} and Faraday rotation measurements \citep{bonafede10,kuchar11}; for instance, rotation measurements for Coma are well fit by $\alpha_{\rm B} \approx 0.3-0.7$ \citep{bonafede10}. In the future it would be interesting to explore other scalings, if for instance this relationship also has temperature dependence \citep{kunz11}. We find that radio surface brightness profiles can be well fit by $\alpha_{\rm B} = 0.3$ for both clusters, but for Perseus $B_{0}=10 \, \mu$G, while for Coma $B_{0}=5 \, \mu$G. We choose a cosmic ray distribution function motivated by cosmological hydrodynamic simulations of galaxy clusters where cosmic rays are accelerated via diffusive shock acceleration \citep{pinzke10}:
\begin{equation}
f_\tr{p}(r,p_\tr{p})=C(r)\sum\limits_i\Delta_i p_\tr{p}^{-\alpha_i}
\end{equation}
\begin{equation}\label{initialdist}
\mathbf{\Delta}=(0.767,0.143,0.0975)\qquad\mathbf{\alpha}=(2.55,2.3,2.15).
\end{equation}
and the normalization
\begin{equation}
C(r)= \frac{(C_{\rm vir}-C_{\rm center})}{1 + \left( \frac{r}{r_{\rm trans}} \right)^{-\beta_{\rm C}}} + C_{\rm center}.
\end{equation}
Note that these simulations do not take into account the effects of cosmic ray streaming. The parameters $C_{\rm vir}, C_{\rm center}, r_{\rm trans}$ are then chosen such that the model radio brightness profile  agrees with fits to observations (equation \eqref{sbfit}). For Perseus, if we define $C(r)=\tilde{C}(r)n_\tr{e}(r)$, then $\tilde{C}_\tr{center}=8.3\ee{-8}, \tilde{C}_\tr{vir}=7.2\ee{-8}$, $r_{\rm trans}=36$ kpc, $\beta_{\rm C}=1.0$. For Coma, $C_{\rm center}=6 \times 10^{-11} \, {\rm cm^{-3}}$, $C_{\rm vir} = 5.2 \times 10^{-11} \, {\rm cm^{-3}}$, $r_{\rm trans}=55$ kpc, $\beta_{\rm C}=1.09$. The initial radio surface brightness profiles derived from these parameters are shown in Fig \ref{perseusplots}a and \ref{comaplots}a.

When we solve equation (\ref{crevol}), the simulation grid has two ghost zones at each end in the radial direction, and two ghost zones at each end in the momentum direction. To set values in the ghost zones, we use ${\rm d\,log} \, f_\tr{p}/{\rm d\,log} \, p=$const in the momentum direction at both the inner and outer boundary, i.e. a power law extrapolation. In the spatial direction, we use ${\rm d\,log} \, f_\tr{p}/{\rm d\,log} \, r=$const at the inner boundary. The outer boundary requires a little more care, since it can fall to extremely low values which result in round-off error; also, if the CR gradient goes to zero at the outer simulation boundary, this artificially suppresses CR streaming. For Perseus, we use ${\rm d\,log} \, f_\tr{p}/{\rm d\,log} \, r=$const at the outer boundary, but subject to the condition that $f_{\rm min} \le f_{i_{\rm max}+1} \le X f_{i_{\rm max}}$, $f_{\rm min} \le f_{i_{\rm max} +2} \le X f_{i_{\rm max}+1}$, where $i_{\rm max}$ is the index of the last active zone, $f_{\rm min}(p) = 10^{-3} f_\tr{p}(r_{\rm max},p, t_{\rm 0})$, and $X=0.98$. For Coma, where the initial profile is already extremely flat, we simply adopt $f_{i_{\rm max}+1} = X f_{i_{\rm max}}$, $f_{i_{\rm max}+2} = X f_{i_{\rm max}+1}$. To conserve CRs during the process of turbulent advection, we also enforce the CR turbulent diffusion flux, defined as ${\mathbf F}_{\rm turb} = \kappa_{\rm turb} \delta^{\alpha/3} \nabla (f_{p} \delta^{-\alpha/3})$ (see equation (\ref{eqn:f_diffusion})), to be zero at both spatial boundaries.

For both Perseus and Coma, we use 1.4 GHz data. Note that for Coma, which is the most well-studied giant radio halo, recent 1.4 GHz and 352 MHz data cannot be reconciled by the classical hadronic model with a power-law spectrum \citep{brown11-coma}, though this conclusion is subject to systematic uncertainties in the zero-point of 1.4 GHz data \citep{zandanel12}. We shall also see that energy dependence in the streaming speed alters the CR distribution function, so that it is no longer a power-law in momentum, potentially solving this problem.

\section{Results}\label{results}

We now show results for a canonical radio mini-halo in a cool core cluster (Perseus), and giant radio halo in a non-cool core cluster (Coma), starting from the initial conditions given in \S\ref{sect:initial_conditions}. We use Perseus to illustrate most of the relevant physics. Unless otherwise noted, all calculations assume $L_{\rm MHD} = 100$kpc, where $L_{\rm MHD}$ is the lengthscale at which $v_{\rm A} = v_{\rm turb}$ (note from $\epsilon = v_{\rm A}^{3}/L_{\rm MHD}$ that smaller values of $L_{\rm MHD}$ correspond to more vigorous turbulence).

\subsection{Perseus Cluster}\label{Perseus}
\begin{figure*}
  \includegraphics[width=7cm,trim=1cm 9cm 5cm 4cm]{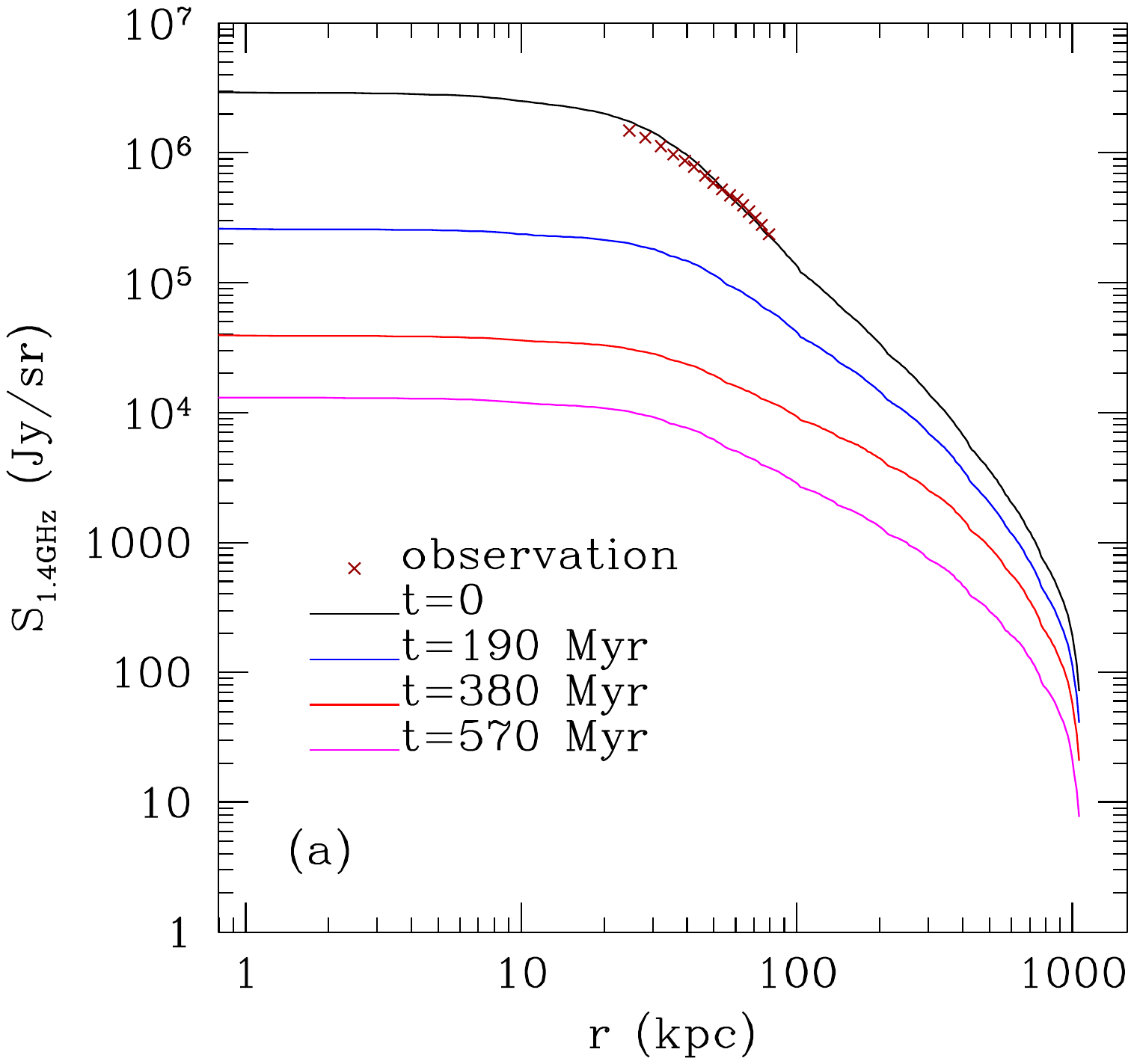}
  \includegraphics[width=7cm,trim=1cm 9cm 5cm 4cm]{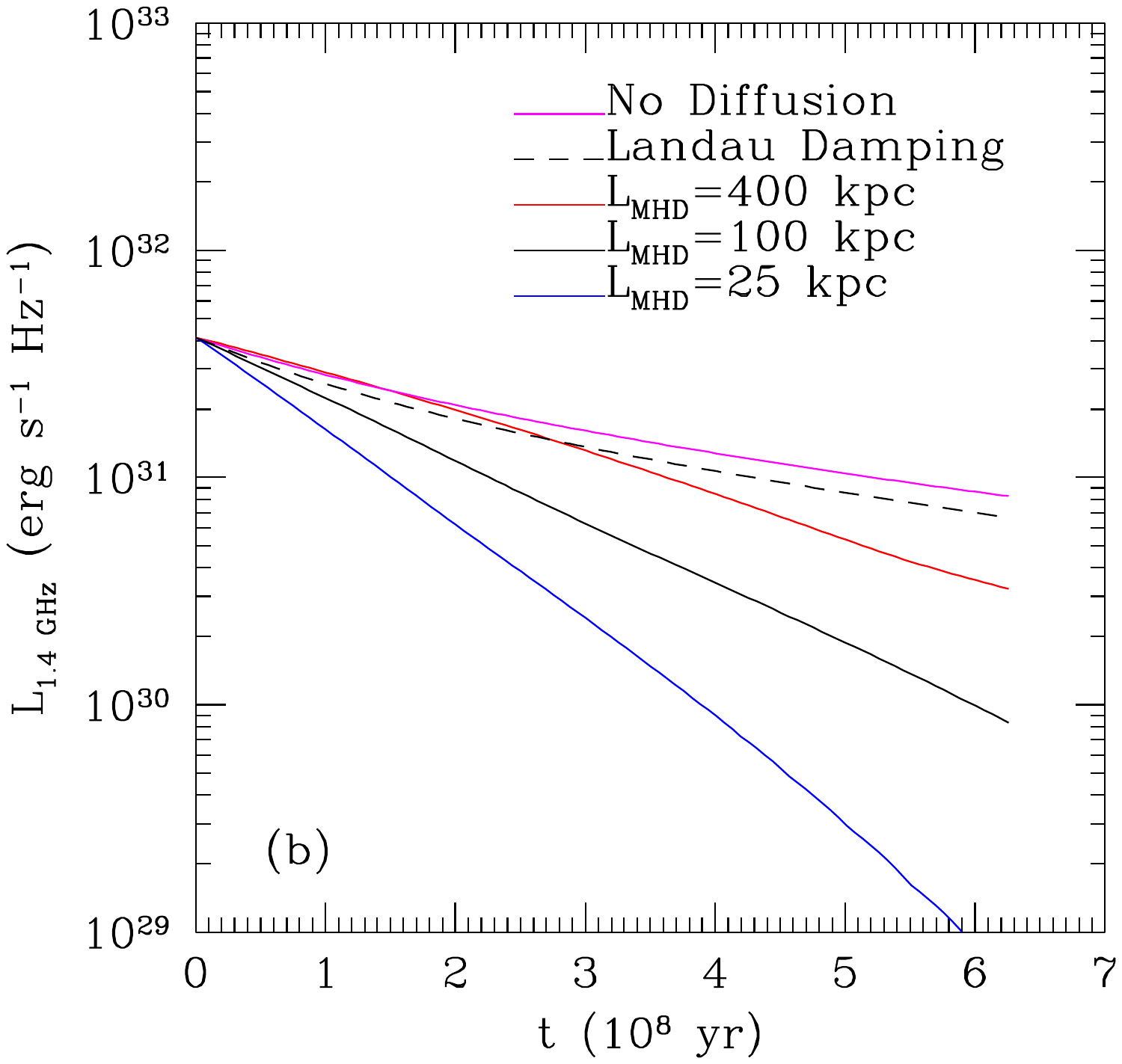}
  \includegraphics[width=7cm,trim=1cm 9cm 5cm 4cm]{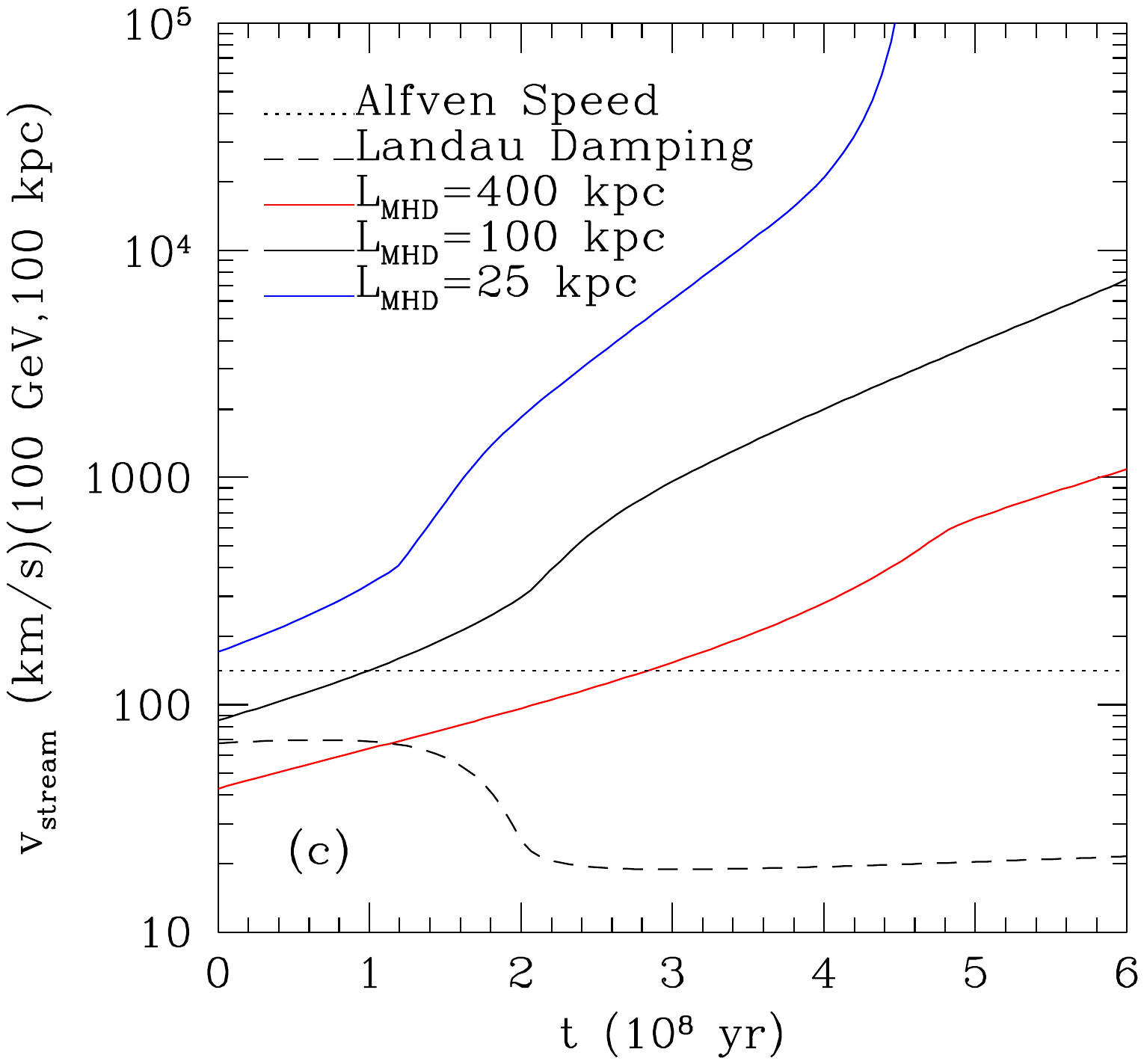}
  \includegraphics[width=7cm,trim=1cm 9cm 5cm 4cm]{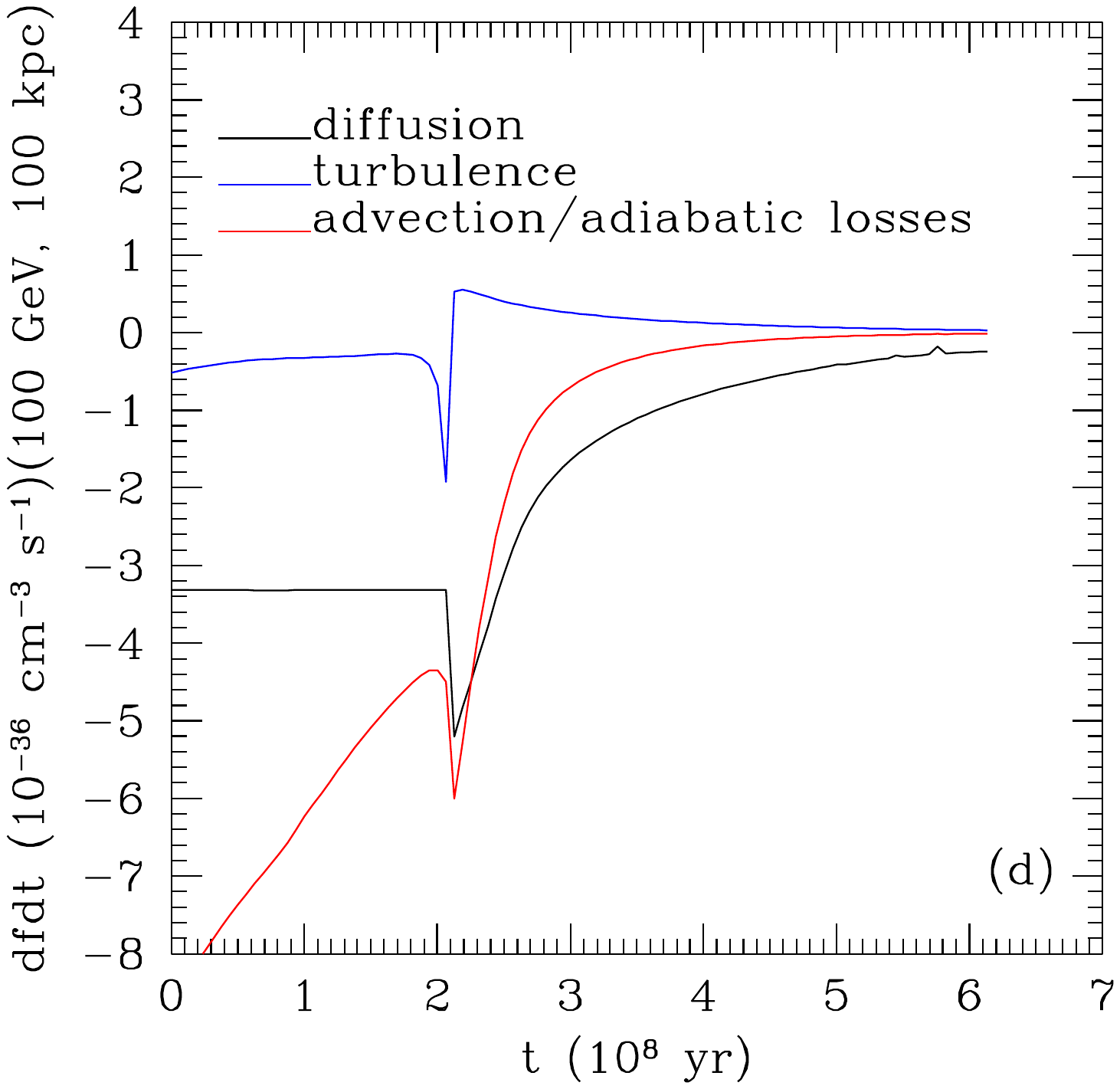}
  \includegraphics[width=7cm,trim=1cm 9cm 5cm 4cm]{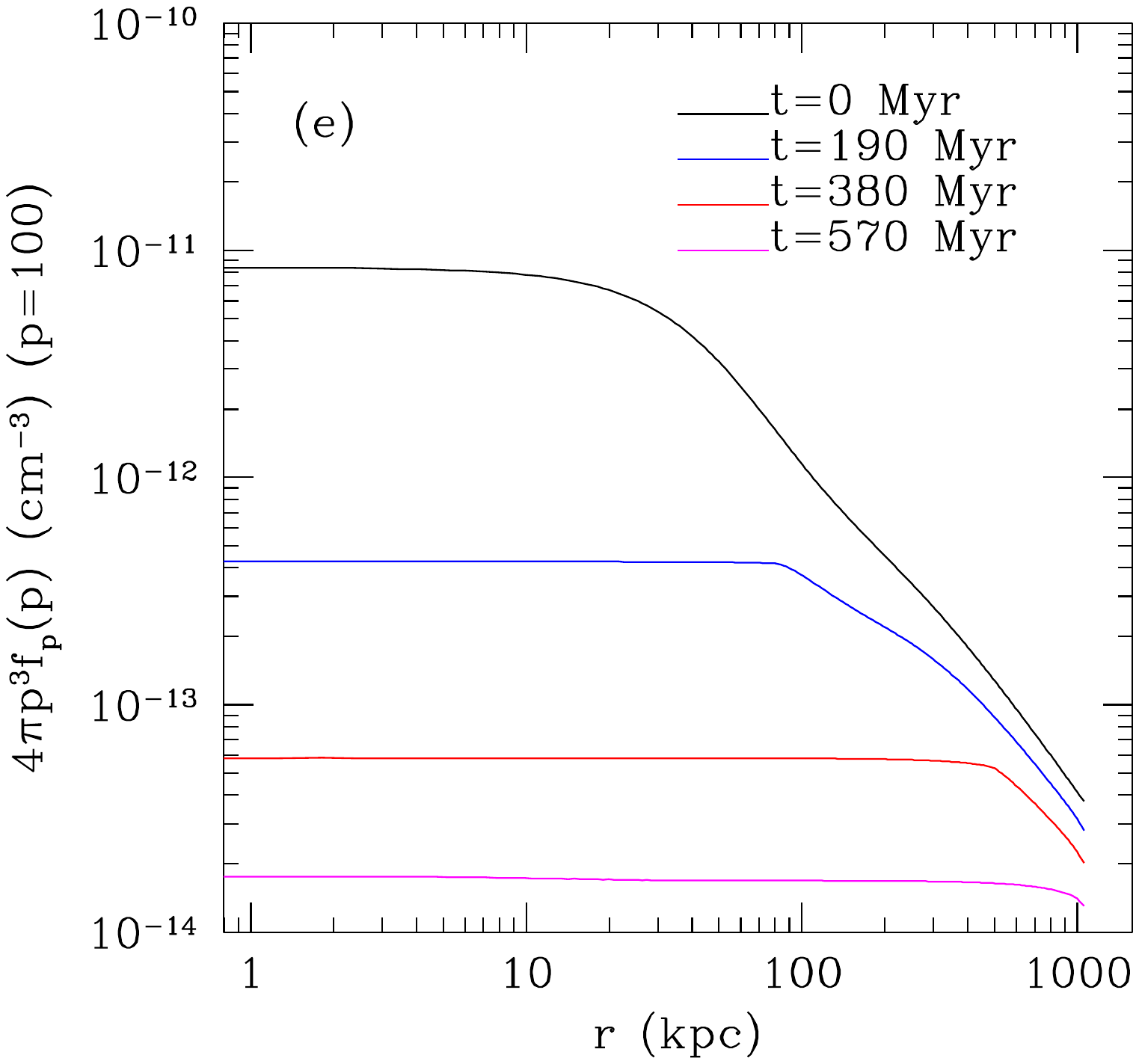}
  \includegraphics[width=7cm,trim=1cm 9cm 5cm 4cm]{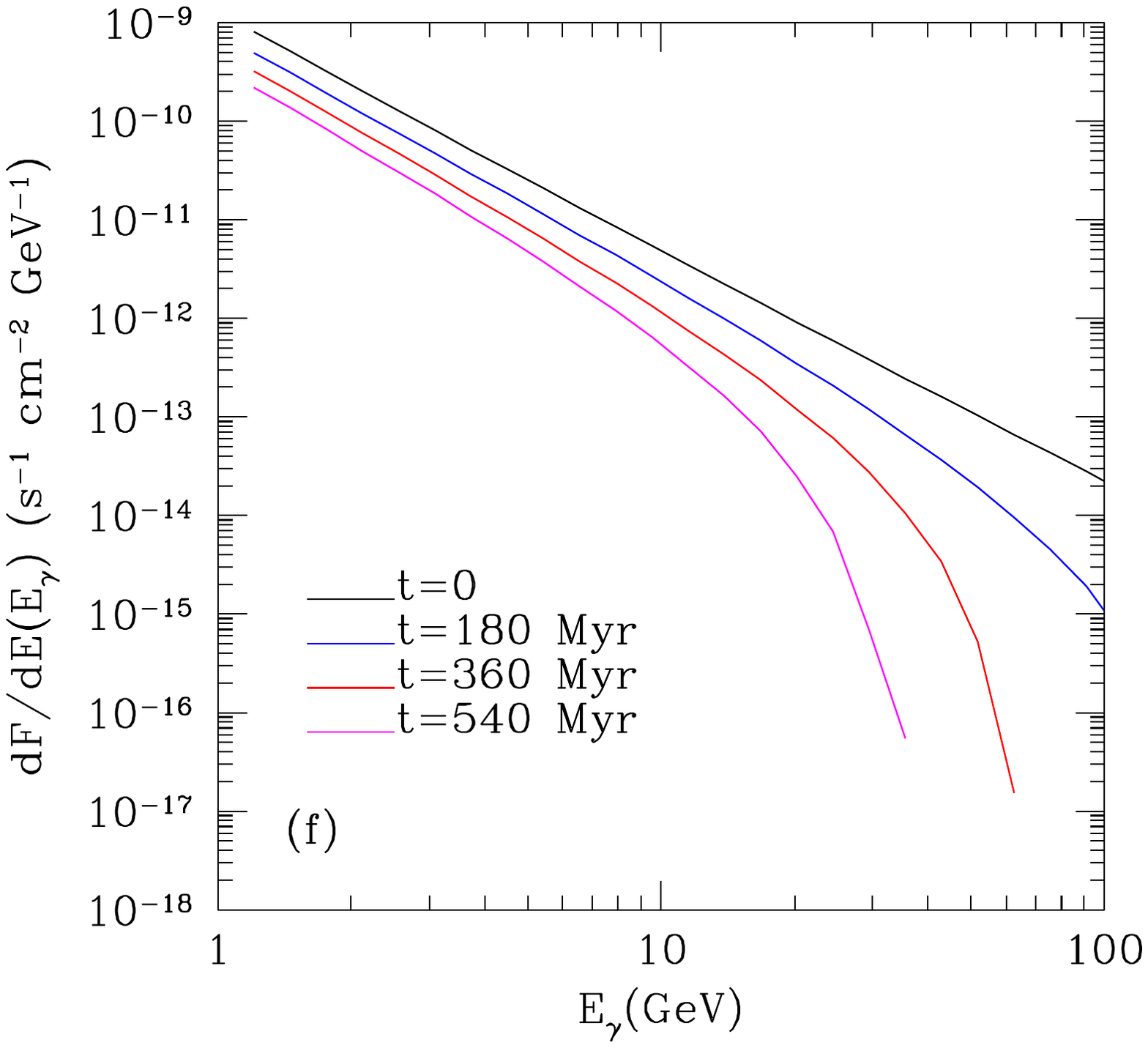}
\caption{Simulation results for the Perseus cluster. \textbf{(a)} Radio surface brightness of Perseus for $L_\tr{MHD}=$ 100 kpc. Observations from \citet{pedlar90}. \textbf{(b)} The time evolution of the Perseus cluster's radio luminosity for different levels of damping. The solid lines show MHD turbulence damping at various strengths. The dashed line shows non-linear Landau damping. \textbf{(c)} Cosmic ray streaming speeds of 100 GeV CRs at a fixed radius of 100 kpc. \textbf{(d)} Different contributions to $\dot{f}_\tr{p}$ for the $L_\tr{MHD}=100$ kpc Perseus simulation. \textbf{(e)} Radial distribution of 100 GeV protons for $L_\tr{MHD}=100$ kpc. \textbf{(f)} Predicted gamma-ray fluxes. Upper limits from observations are at higher energies than those plotted here.}\label{perseusplots}
\end{figure*}

The initial conditions for Perseus correspond to a CR profile where the ratio of CR energy to the thermal gas energy is almost constant throughout the cluster, decreasing slightly in the outskirts. In Fig \ref{perseusplots}a we compare radio emission from our initial conditions to surface brightness observations at 1.4 GHz from \cite{pedlar90}. Note that the observations only span a limited radial range (which produces $\sim 1/2$ of the total radio luminosity in our model). The normalization of the profile falls substantially in several hundred Myr, while its shape does not evolve significantly\footnote{This is mostly due to projection effects; note that the CR radial profile {\it does} evolve significantly; see Fig. \ref{perseusplots}e.}. Fig \ref{perseusplots}b shows the evolution of the 1.4 GHz radio luminosity with time, and how it depends on the strength and nature of loss processes. The fall in luminosity is exponential, on a characteristic $\sim 10^{8}$ yr timescale. For our fiducial $L_{\rm MHD} =100$ kpc simulation, $L_{\rm 1.4 GHz}$ falls by an order of magnitude in several hundred Myr; the decrease is faster for smaller values of $L_{\rm MHD}$, which corresponds to strong damping. If only non-linear Landau damping operates, the decline in luminosity is very slow, and insufficient to turn off radio halos. We also show how $L_{\rm 1.4 GHz}$ evolves if we ignore diffusion (i.e., the no damping limit) or adiabatic losses in equation (\ref{crevol}). %Interestingly, no one process dominates; radio halo turn-off is much more efficient if both processes act in tandem. We will return to this point later.

The streaming speeds relative to the wave frame for 100 GeV CRps at a radius of 100 kpc are shown in figure \ref{perseusplots}c, for different values of $L_{\rm MHD}$ and if only non-linear Landau damping dominates. We see that even if the streaming speeds start out slow, they can quickly become super-Alfv\'enic as the CR density drops. This non-linear behavior, which is due to the unusual $\kappa \propto 1/\nabla f$ scaling of the diffusion coefficient for turbulent damping, allows very fast streaming. It is not seen if only non-linear Landau damping operates; in that case, $v_{\rm D} \sim {\mathcal O}(v_{\rm A}$) at all times. Note from equation (\ref{eqn:v_D_intermediate}) that $v_{\rm D}-v_{\rm A} \sim (\lambda/3L_{\rm z})c$. Thus, as $v_{\rm D} \rightarrow c$, $\lambda \rightarrow L_{z}$, and our equations break down, as the CRs can no longer be described by a distribution function. Instead, a fully kinetic approach is needed. This limitation is relatively unimportant since by this stage CRs are no longer self-confined but stream freely along field lines; thus, turn-off is extremely rapid.

The individual contributions to $\dot{f}_\tr{p}$ at 100 GeV are shown in Figure \ref{perseusplots}d in the $L_\tr{MHD}=100$ kpc case. The values are taken at a fixed radius of 100 kpc, and displayed as a function of time. Interestingly, no one process dominates (and we have verified that adiabatic and diffusive losses acting in tandem are much more effective than either process alone). Initially, adiabatic losses dominate, although they decrease continuously with time. This is to be expected, since the adiabatic loss term is proportional to $f_\tr{p}$, and decreases as $f_\tr{p}$ falls. On the other hand, the diffusion loss term from turbulent damping is independent of $f_\tr{p}$ (equation \eqref{eqn:diffusion_turb}), and thus independent of time as long as there is a spatial gradient. At $t \sim 190$ Myr, the profile at $r \sim 100$ kpc flattens (see Fig. \ref{perseusplots}e), and all terms plummet, although the adiabatic loss term falls most drastically. In \S\ref{sect:analytic}, we explore the nature of this change when the profile flattens: inside the flat core, $\dot{f_\tr{p}}$ changes since it is determined solely by the flux at the outer boundary of the core. From this plot, we can see why adiabatic and diffusive losses in tandem are much more efficient that either alone: adiabatic losses are much more effective in the early stages when the profile is centrally peaked, while diffusive losses are more effective once the profile flattens. We also see that inward turbulent advection is non-negligible but subdominant. It also changes once the region becomes incorporated inside the flat core (since once again only the flux through the core boundary matters at that point).

Fig \ref{perseusplots}e shows $4 \pi p^{3} f_{\rm p}$ (i.e., CR density) for 100 GeV CRs. As previously discussed, the CR density profile develops a flat core, which expands in size at roughly the streaming speed. Meanwhile, the normalization of the CR profile falls continuously, even for the flat portion. The end result is a profile in which the CR profile has completely flattened and fallen by several orders of magnitude by the end of the simulation.

Finally, the expected $\gamma$-ray flux as calculated from equation (\ref{eqn:gammaflux}) is shown in Fig \ref{perseusplots}f. Observed upper limits are also shown; note that our initial conditions are consistent with these upper limits. Due to the finite momentum grid $p_{\rm p} \le 5000$, our calculations are only accurate in the range $E_{\gamma} \lsim 200$ GeV, although the high energy CRs stream so quickly that the CR transport equation quickly breaks down, in any case. The gamma-ray fluxes decline extremely rapidly with time, with the decline being much sharper at higher energies, due to the fact that higher energy CRs stream faster. The upshot is that at the $E_{\gamma} \sim 0.3-1$ TeV ($E_{\rm CR} \sim 3-10$ TeV) energies probed by imaging air Cherenkov telescopes (MAGIC, HESS, VERITAS), the decline in gamma-ray luminosity is very rapid. Any detection of gamma-ray emission at these energies, where a source is not immediately apparent (suggesting that it is long-lived), would strongly disfavor the model of CR streaming presented here. However, the $E_{\gamma} \sim 0.1-3$ GeV ($E_{\rm CR} \sim 1-30$ GeV) energies probed by Fermi correspond to CRs which stream and turn off gamma-ray emission more slowly. The latter is thus a more robust measure of the cluster's CR injection history. Note that since $\langle E_{\gamma} \rangle \sim 1/8 \langle E_{\rm CR} \rangle$, gamma-ray emission at $E_{\gamma} \sim 10$ GeV corresponds to the $E_{\rm CR} \sim 100$ GeV CRs relevant for $\sim$GHz radio emission, and declines by a similar amount.

\begin{figure}
\includegraphics[width=8cm,trim=2cm 10cm 4cm 4cm]{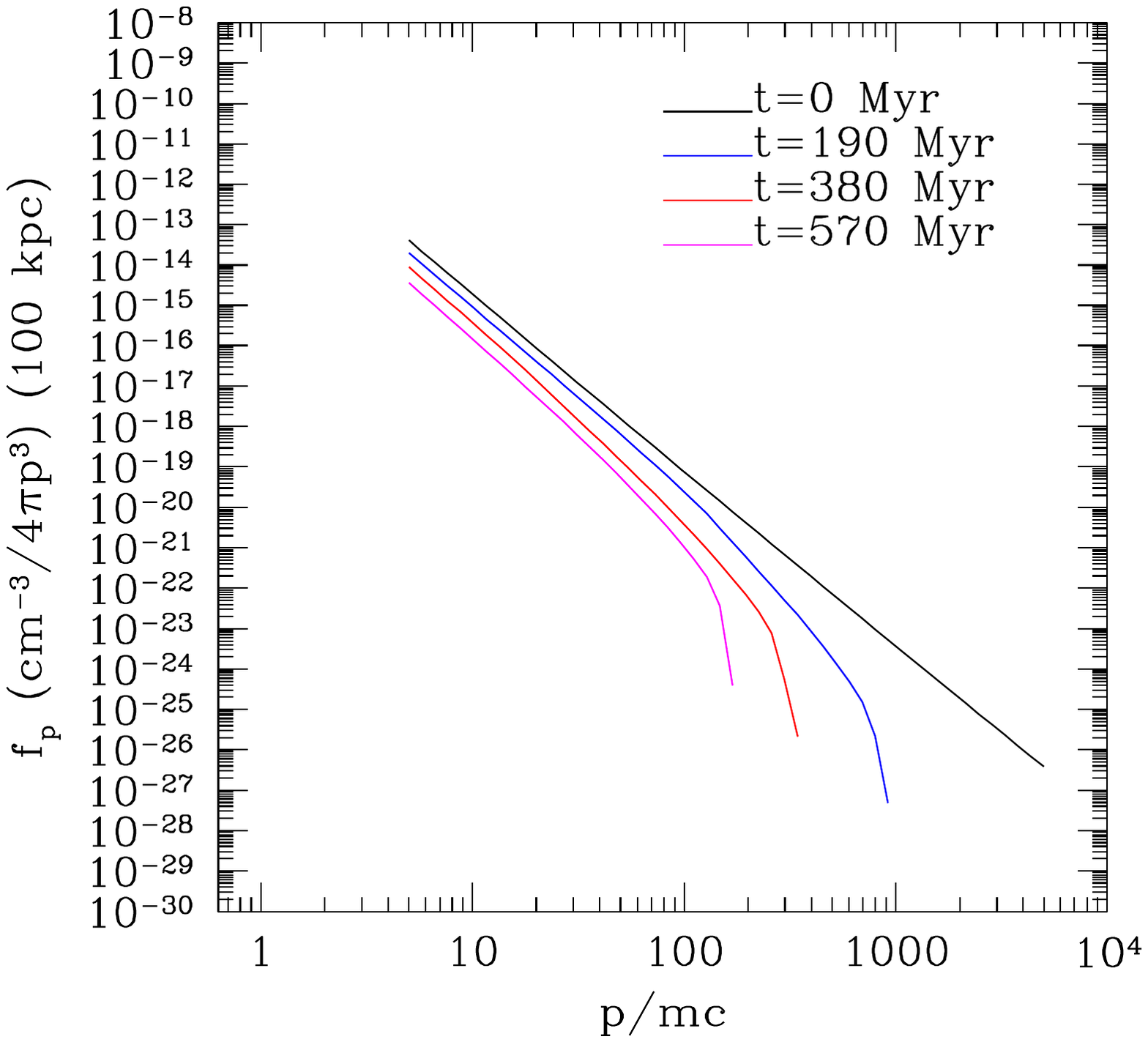}
\caption{CR distribution function versus momentum at a fixed radius of 100 kpc. The dropoff time scales with energy, leading to the spectral steepening discussed above.}
\label{fp0}
\end{figure}

By the same token, the energy dependence of CR streaming implies that radio luminosity turns off more slowly at lower frequencies. We show this in Fig \ref{rlumw2}. We can also see this in figure \ref{fp0} which plots the distribution function versus momentum. The higher energy CRps drop in density much faster than lower energy CRps. The corresponding high energy synchrotron emission then also drops faster. This behavior could explain radio halos such as Abell 521, which is detected at 240, 325 and 610 MHz, but not at 1.4 GHz, implying a cutoff or strong spectral curvature at high frequencies \citep{brunetti08}. We therefore predict that at the low frequencies probed by LOFAR, radio halos should be significantly more abundant.

\subsection{Coma Cluster}

\begin{figure*}
  \includegraphics[width=7cm,trim=1cm 9cm 5cm 4cm]{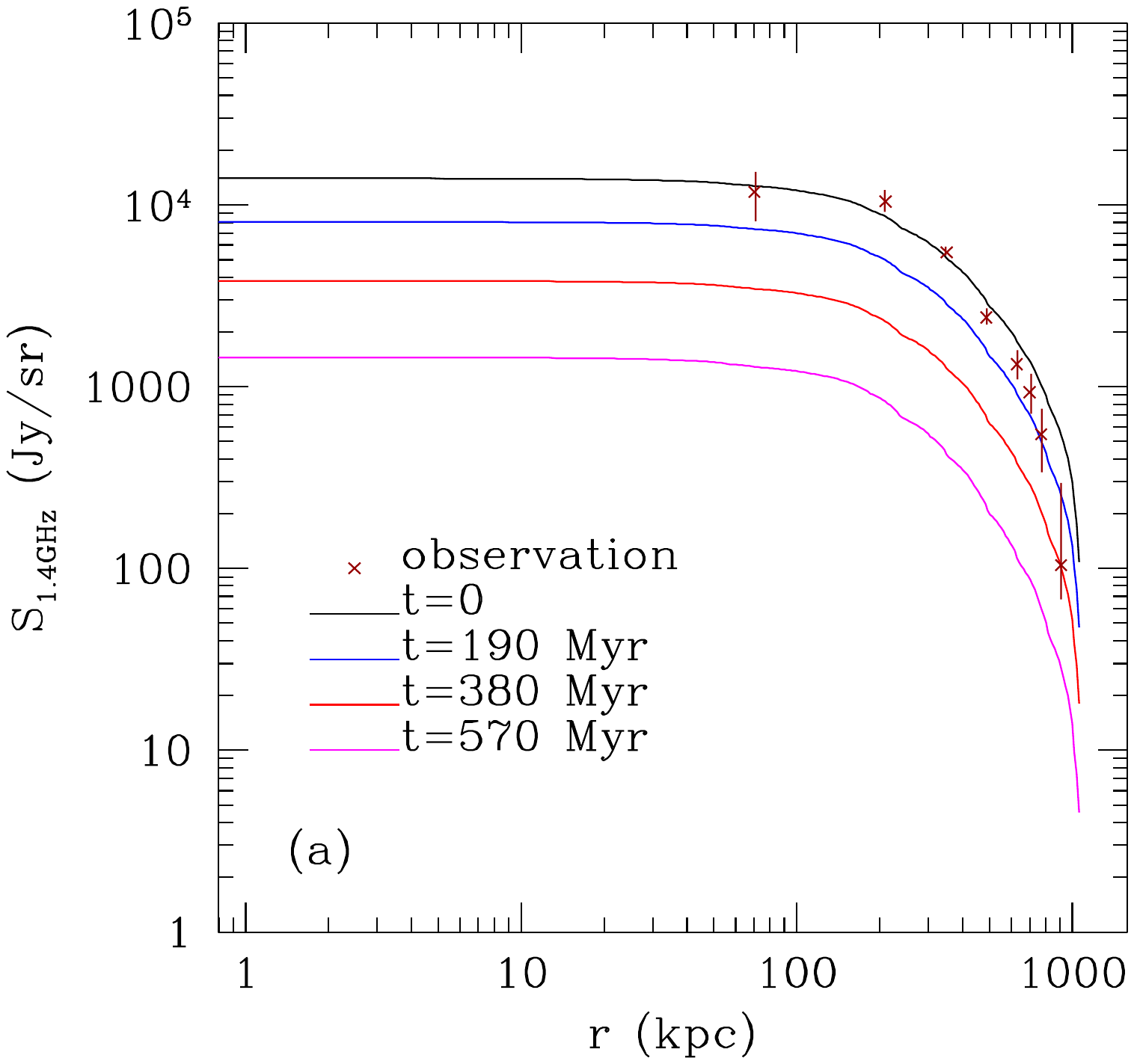}
  \includegraphics[width=7cm,trim=1cm 9cm 5cm 4cm]{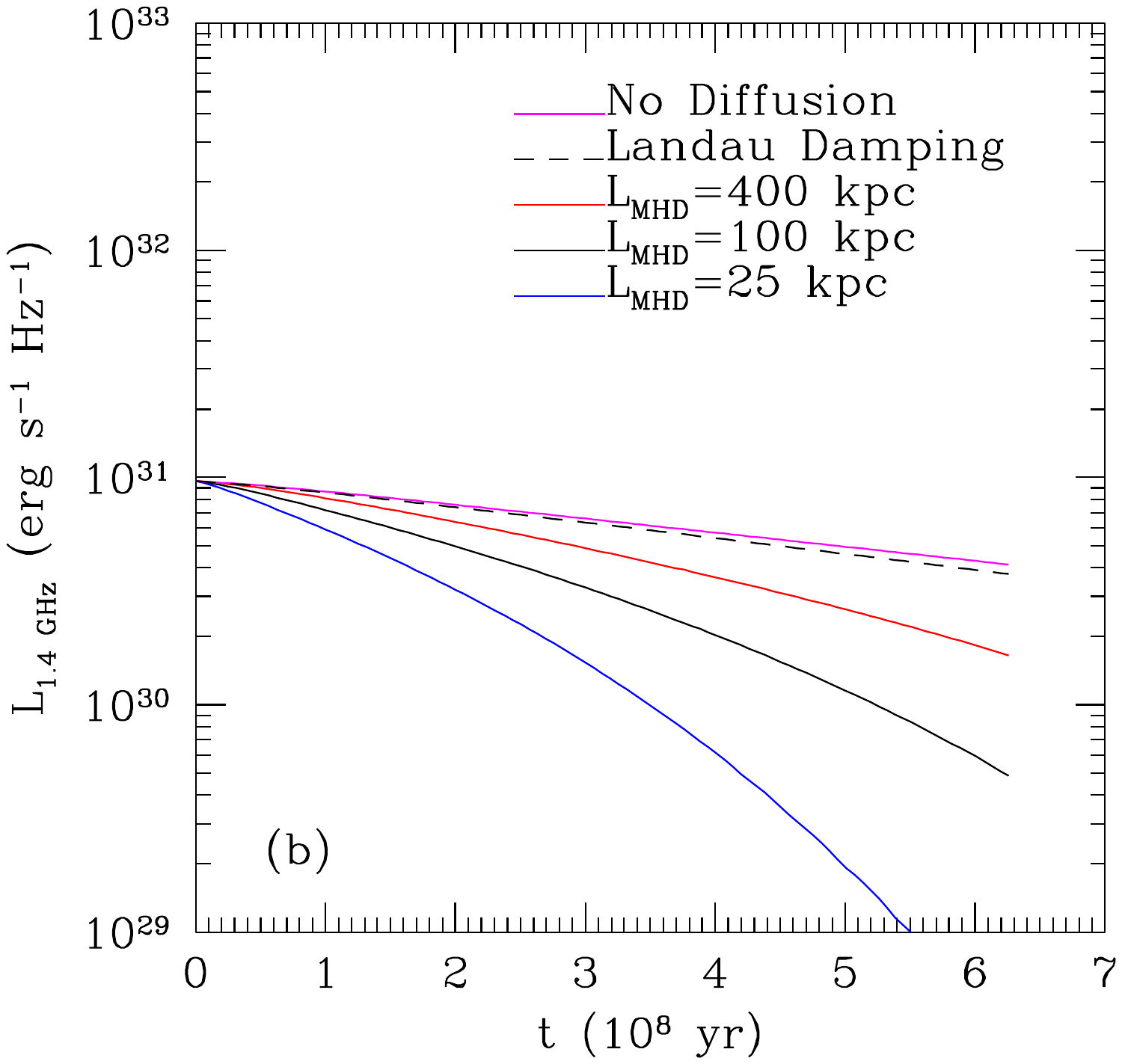}
  \includegraphics[width=7cm,trim=1cm 9cm 5cm 4cm]{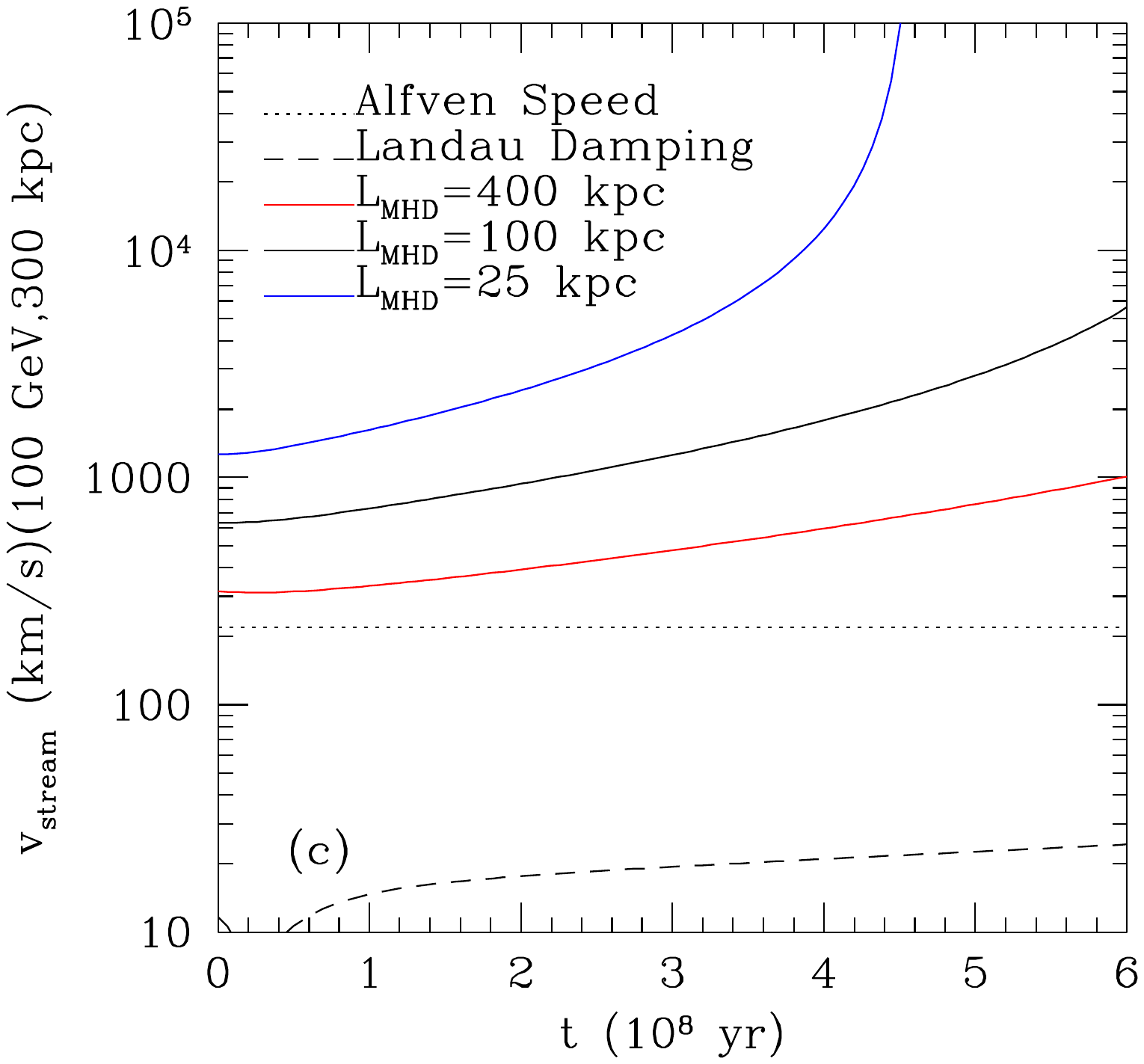}
  \includegraphics[width=7cm,trim=1cm 9cm 5cm 4cm]{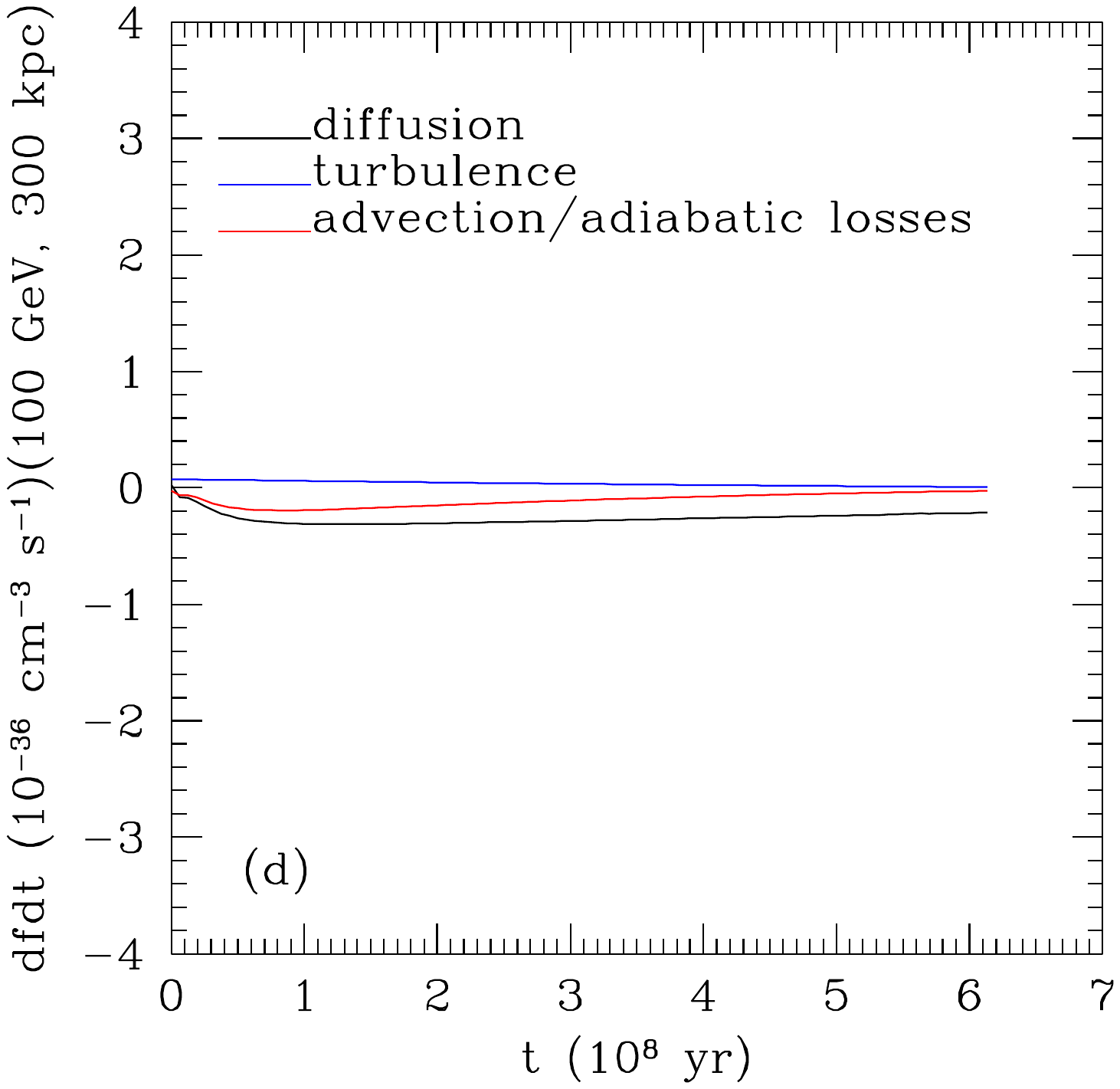}
  \includegraphics[width=7cm,trim=1cm 9cm 5cm 4cm]{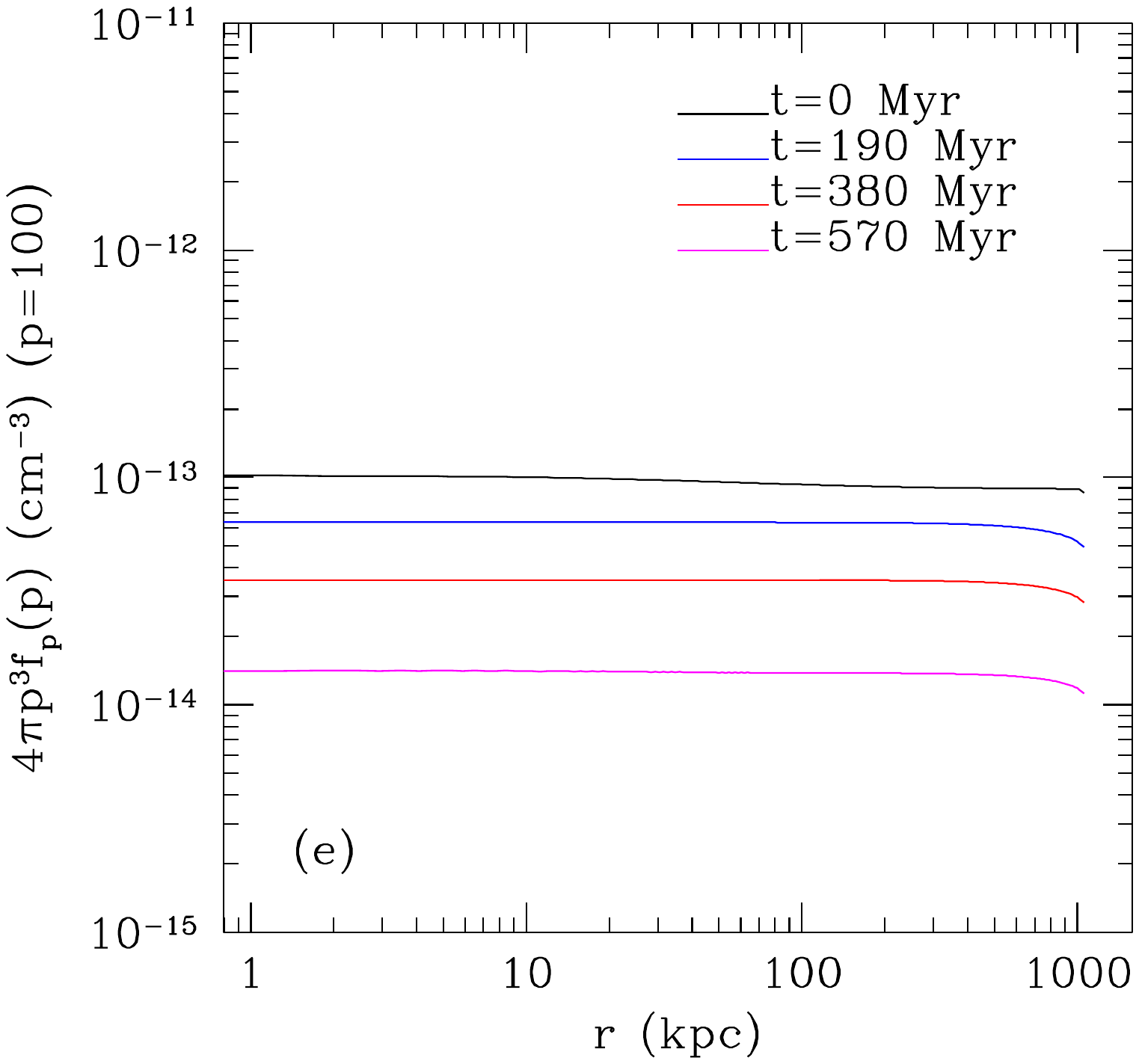}
  \includegraphics[width=7cm,trim=1cm 9cm 5cm 4cm]{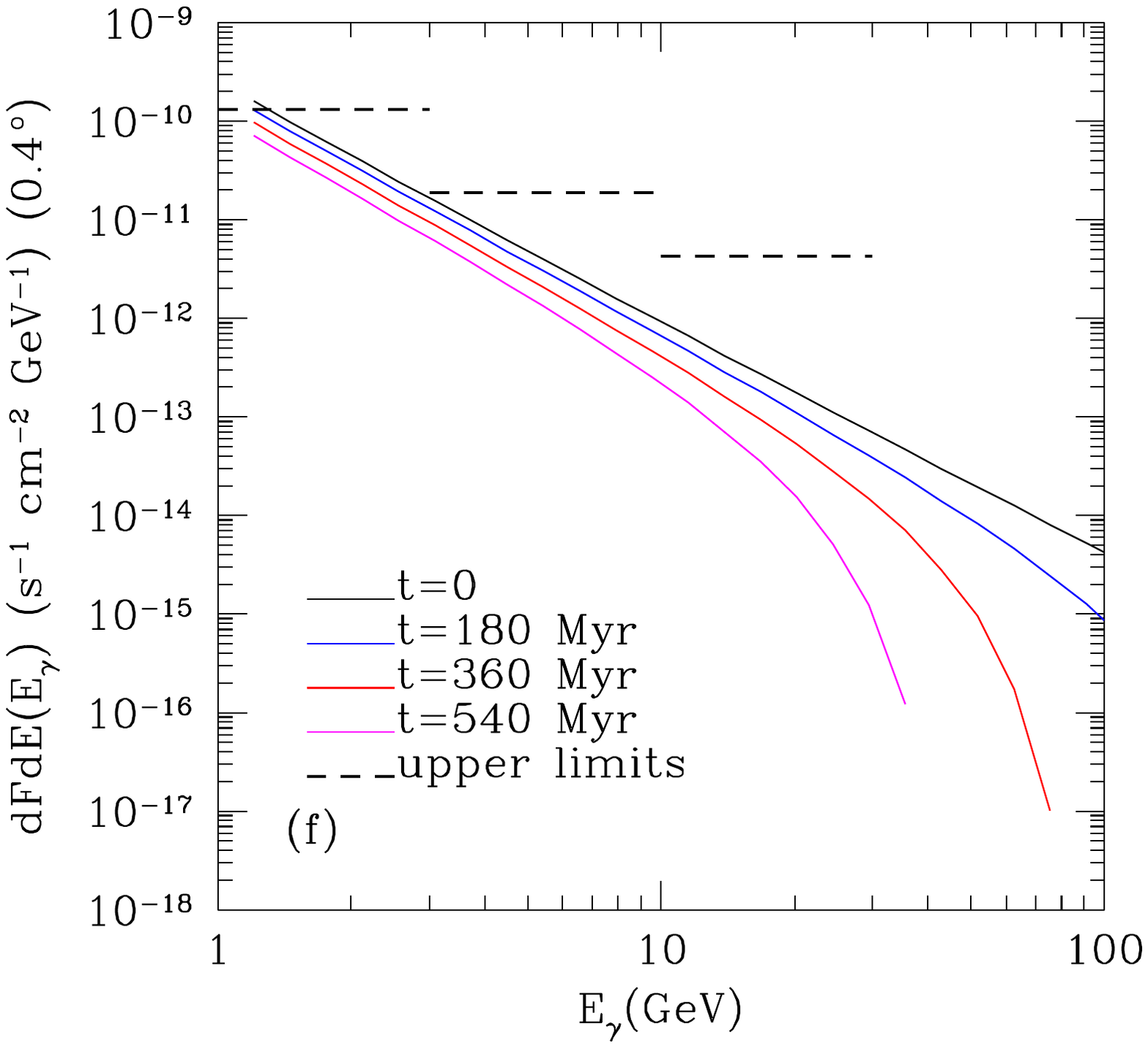}
\caption{Simulation results for the Coma cluster. \textbf{(a)} Radio surface brightness of Coma for $L_\tr{MHD}=$ 100 kpc. Observations from \citet{deiss97}. \textbf{(b)} The time evolution of the Coma cluster's radio luminosity for different levels of damping. The solid lines show MHD turbulence damping at various strengths. The dashed line shows non-linear Landau damping. \textbf{(c)} Cosmic ray streaming speeds of 100 GeV CRs at a fixed radius of 300 kpc. \textbf{(d)} Different contributions to $\dot{f}_\tr{p}$ for the $L_\tr{MHD}=100$ kpc simulation. \textbf{(e)} Radial distribution of 100 GeV protons for $L_\tr{MHD}=100$ kpc. \textbf{(f)} Predicted gamma-ray fluxes. Upper limits are taken from \citep{fermi12} with $\alpha=2.5$.}
\label{comaplots}
\end{figure*}

We now turn to Coma, a prototypical giant radio halo. We focus on the differences with our previous example, Perseus. Due to the flat and extended observed surface brightness profile, which extends out to 1 Mpc (Fig \ref{comaplots}a, observations from \citet{deiss97}), the inferred CRp distribution is much flatter. Indeed, for the B-field we have assumed, $B \propto \rho^{\alpha_{\rm B}}$, $\alpha_{\rm B} \approx 0.3$ (which is consistent with rotation measure observations \citep{bonafede10}), the radio profile can be fit by a nearly flat CRp density (Fig. \ref{comaplots}e). This large flat inferred profile is suggestive  that extensive streaming has already taken place. Coma thus presents an interesting challenge, to see if a significant decline in luminosity is possible despite the absence of significant CR gradients except a small one at the outer boundary. We emphasize once again that for turbulent wave damping, our solutions are independent of the magnitude of the CR gradient $\nabla f$. Our solutions depend only on where $\nabla f$ is non-zero, and its sign.

We show our results in Fig. \ref{comaplots}. All figures are analogs of those for Perseus in Fig. \ref{perseusplots} (except we adopt r=300 kpc as our fiducial radius when displaying time-varying quantities---due to the much larger extent of the Coma radio halo, this is a more representative radius), and we again adopt $L_{\rm MHD}=100$ kpc for our fiducial model. In Fig \ref{comaplots}a, we see that the surface brightness falls in normalization, but does not significantly change shape, as for Perseus. The decline in $L_{\rm 1.4 GHz}$ is slower than for Perseus, although $L_{\rm 1.4 GHz}$ is still down by an order of magnitude after $\sim 600$ Myr for the fiducial model, and declines more quickly with more vigorous turbulence as expected. Non-linear Landau damping alone produces a slow decline. While streaming is super-Alfv\'enic (Fig \ref{comaplots}c), it does not `run away' with time as quickly as for Perseus. The acceleration of CR streaming is tied to the decline of the CR density, which is slower in this case.

Since the flat region encompasses the entire cluster at the outset, there is no transition in energy loss regimes as for Perseus (where the profile gradually flattens). Instead, loss rates vary mildly with time (Fig \ref{comaplots}d), with diffusive losses always more important than adiabatic losses, which are essentially negligible. This can be understood from the fact that adiabatic flux at the outer boundary scales with $f_{\rm p}(R_{\rm max})$, which is small (see \S\ref{sect:analytic} for more discussion), whereas the diffusive flux is independent of $f_{\rm p}(R_{\rm max})$. The flat CR density profile simply decreases in normalization with time (Fig \ref{comaplots}e). Similar to Perseus, Coma's gamma-ray flux declines quickly with time, particularly at the high energies associated with imaging air Cherenkov telescopes.

\begin{figure}
\includegraphics[width=8.5cm,trim=2cm 10cm 4cm 4cm]{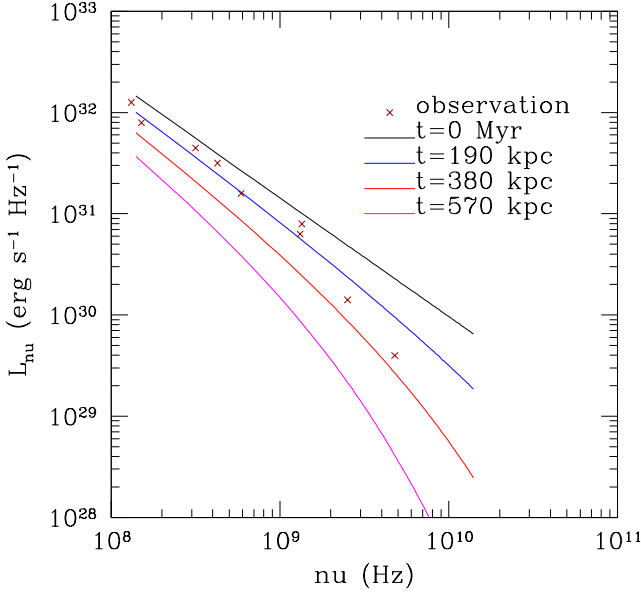}
\caption{Luminosity as a function of energy for the Coma simulation including observations from \citep{brunetti12}. The momentum dependence of the streaming speed leads to a spectral steepening very similar to observation.}
\label{rlumw}
\end{figure}

\begin{figure}
\includegraphics[width=8.5cm,trim=2cm 10cm 4cm 4cm]{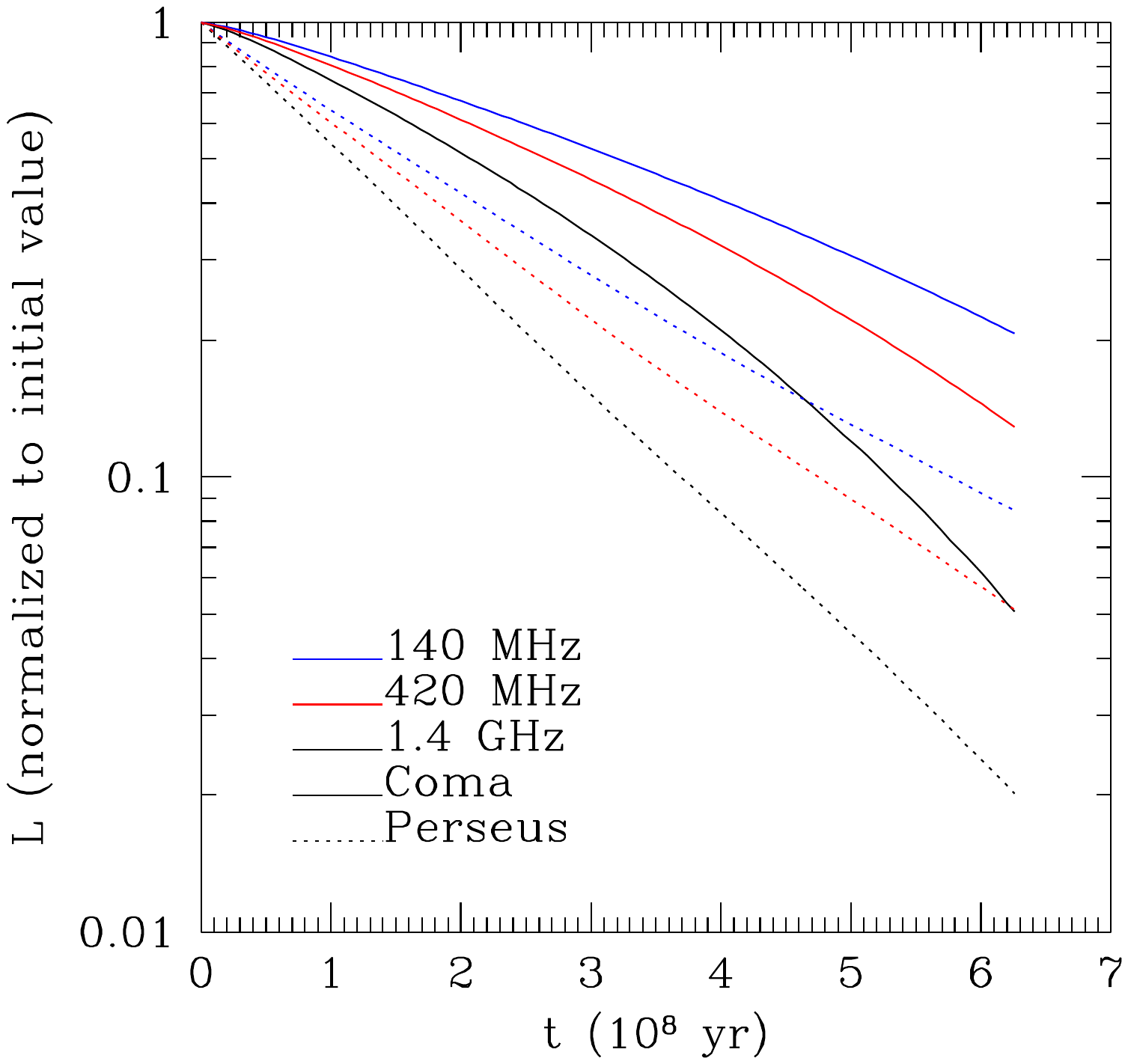}
\caption{Luminosity dropoff in Coma for different frequencies. High energy CRs stream more quickly, so the higher frequency signals drop faster.}
\label{rlumw2}
\end{figure}

Spectral steepening in Coma's radio emission has been seen in multi-frequency observations \citep{brunetti12}, a feature which occurs naturally in our models due to the energy dependence of CR streaming. We show this in Fig \ref{rlumw}; spectral steepening very similar to that observed arises. Given the flat inferred profile of Coma, which suggests that substantial streaming has already taken place, this raises the possibility that a power-law population with a slightly higher normalization was transformed by streaming into the curved population we see today. %As high-energy CRs stream ahead of low energy CRs, they encounter lower density gas and consequently produce secondary electrons via hadronic processes at a lower rate.  These electrons in turn emit less synchrotron  radiation due to the weaker B-fields at large radii. The first factor is well-constrained by observed density profiles, but the variation of B-field strength with radius is less well-constrained. {\textbf} If we assume a steeper scaling $B \propto \rho^{\alpha_{\rm B}}$ with $\alpha_{\rm B} \approx 0.8$ the rate of turn-off and development of spectral steepening will be faster.

%%{\textbf An important caveat to our results is our}

We have chosen a rather extreme case of a completely flat profile, to illustrate that radio halo turn-off is still possible in this case. The observational data also permit an initial CR profile that is less flat. Using the same B-field, we can still reproduce the observations very well with a profile that has a mild central peak. Since we now have a significant density gradient, the radio luminosity can drop off faster in the beginning, although the overall evolution is qualitatively the same as before. The streaming speeds ramp up faster than in the flat profile fit.

We have assumed a magnetic field profile $B \propto \rho^{\alpha_{\rm B}}$, with $\alpha_{\rm B} = 0.3$. This choice assumes the B-field is in rough equipartition with turbulence, as discussed in \S\ref{sect:turb}, agrees with Faraday rotation measures, and enables us to reproduce the observed surface brightness distribution. However, the Faraday rotation measurements are consistent with a range of values $\alpha_{\rm B} \sim 0.4-0.7$ at 1$\sigma$ \citep{bonafede10}. A steeper scaling of $B$ with density implies lower B-fields at the cluster outskirts; in this case the rate of turn-off and development of spectral steepening will be slower. While lowering B decreases the Alfv\'en speed, the dominant effect is a decrease in the CR flux $F$ due to streaming, which scales as $B^{3/2}/L_{\rm MHD}^{1/2} $ (equation \eqref{eqn:diffusion_turb}). The rate of CR streaming is set by the minimum value of this flux, which generally occurs at the cluster outskirts. As we shall we shall see in \S\ref{sect:analytic}, in this regime we can approximate $\dot{f_\tr{p}} \approx 3F(R_{f},p)/R_{f}$. Specializing to our model for Coma ($4\pi p^3 f_\tr{p}(t=0)\approx10^{-13}\tr{ cm}^{-3}$ for $p=100$, $n_\tr{e}(0)/n_\tr{e}(1\tr{ Mpc})\approx17$, and $B_0=5\ \mu\tr{G}$):
\begin{equation}
t_{\rm off} \sim \frac{f_\tr{p}}{\dot{f_\tr{p}}}\bigg{|}_{1\tr{ Mpc}} \sim 370 \, {\rm Myr} \,L_{\rm MHD,100}^{1/2} 70^{\alpha_B-0.5}
\end{equation}
Thus, a steeper scaling $\alpha_{\rm B} = 0.7$ would not permit turn off on an acceptably short timescale for a very flat profile (it could still be possible for a less flat profile, as above, but note that the observations cannot be fit well by a centrally peaked CR profile with this steep B-field scaling). Note that radio relic measurements are consistent with strong, $\sim \mu$G fields at the cluster outskirts \citep{feretti12}; in addition, the very strong turbulence at the cluster outskirts could be consistent with lower values of $L_{\rm MHD}$ than the constant value we have assumed. As we reiterate in the Conclusions, complex issues regarding magnetic field topology and strength are best further explored with 3D MHD simulations.

\section{Analytic Expressions}\label{sect:analytic}

In certain limiting cases, the evolution of the CR population can be derived analytically. These solutions serve two purposes: they serve as tests of our numerical code, particularly the regularization scheme (\S\ref{sect:stability}), and they also give physical insight into the behaviour of our solutions, and the circumstances under which particular processes dominate.

In the absence of sources ${Q}$ and ignoring the negligible Coulomb and hadronic losses, we can write the CR transport equation (\ref{crevol}) as:
\begin{equation}
\frac{{\rm D}f_\tr{p}}{{\rm D}t} \approx \frac{\partial f_\tr{p}}{\partial t} \approx - \nabla \cdot {\mathbf F}
\label{eqn:crevol_approx}
\end{equation}
where the total CR flux ${\mathbf F}= {\mathbf F}_{\rm adia} + {\mathbf F}_{\rm str} + {\mathbf F}_{\rm turb}$, is made up of the fluxes due to adiabatic losses in the wave frame, streaming relative to the wave frame, and turbulent advection respectively. We have approximated the Lagrangian derivative by the Eulerian derivative ${{\rm D}f_\tr{p}}/{{\rm D}t} \approx {\partial f_\tr{p}}/{\partial t}$, since ${\mathbf v}_{\rm A} \cdot \nabla f_\tr{p}$ is initially small and becomes increasingly negligible as the profile flattens.

As we have seen, the CR profile generally develops a flat inner core within some radius $R_{f}$, outside of which it declines. The flat core stems from the fact that while $\nabla \cdot {\mathbf F}$ increases inward\footnote{This condition holds as long as ${\mathbf F}$ increases more slowly than $r$. In our case, the dominant fluxes $F_{\rm stream} \propto B^{3/2} \propto \rho^{3\alpha_{\rm B}/2}$ (equation \eqref{eqn:F_stream}) clearly increases inward, and $F_\tr{adia} \propto f {\mathbf v}_{\rm A}$ is at most flat or increases inward. Thus, $|\nabla \cdot {\mathbf F}|$ clearly increases inward.},  an inverted CR profile cannot develop, since CRs cannot stream up a gradient. Thus, a flat core develops, while its normalization and radius $R_{f}$ evolves due to the net flux of CRs from its outer boundary. In particular, if we set ${\mathbf F}=F \hat{\mathbf r}$ and integrate equation \eqref{eqn:crevol_approx} over the volume of the flat region, we obtain:
\[
\frac{4}{3}\pi R_{f}^3\dot{f}_\tr{p}(R_{f},p)=-4\pi R^2F (R_{f},p)
\]
\begin{equation}
\dot{f}_\tr{p}(R_{f},p)=-\frac{3F (R_{f},p)}{R}
\end{equation}
where we have used the fact that $\dot{f_\tr{p}}(r,p,t)$ is independent of $r$ for $r<R_{f}$, and the divergence theorem. The evolution of the entire profile can then be described by
\begin{equation}
\dot{f}_\tr{p}(r,p,t)=
\begin{cases}
-\frac{3F(R_f,p,t)}{R_f(p,t)},\qquad r<R_{f}(p,t)\\
-\nabla\cdot\mathbf{F}(r,p,t),\qquad r>R_{f}(p,t)
\end{cases}
\label{eqn:analytic}
\end{equation}
where the ``flatness front'' $R_{f}(p,t)$ is determined from
$f_\tr{p}(0,p,t)=f_\tr{p}(R_{f},p,t)$, or:
\begin{eqnarray}
f_\tr{p}(0,p,0)-\int_0^t\frac{3F(R_\tr{f}(p,t'),p,t')}{R_{\rm f}(p,t')}\tr{d}t'=
f_\tr{p}(R_{f},p,0) \nonumber \\ -\int_{0}^{t} \nabla\cdot\mathbf{F}(R_\tr{f}(p,t'),p,t') dt'
\label{eqn:analytic_front}
\end{eqnarray}

As we have seen, ${\mathbf F}_\tr{stream}$ and ${\mathbf F}_\tr{adia}$ are the most important fluxes, while ${\mathbf F}_\tr{turb}$ is subdominant. Let us now consider the limiting cases when only one is at play.

{\bf Cosmic-Ray streaming only.} We have:
\begin{equation}
\mathbf{F}_\tr{stream}=\frac{\Gamma_{\rm D} B^2\hat{r}}{4\pi^3p^3m\Omega_0v_\tr{A}}=F_\tr{stream}\hat{r}
\label{eqn:F_stream}
\end{equation}
Since $F_\tr{stream}$ is independent of $f_\tr{p}$ and depends only on plasma parameters (specifically, the B-field, turbulence and density profiles), in our model where the gas properties are time-steady (and thus in hydrostatic and thermal equilibrium), $F_\tr{stream}(r,p)$ is independent of time. Thus, $A(r,p) \equiv \nabla \cdot \mathbf{F}_\tr{stream}(r,p)$ is also time-independent, and we have for $r>R_{f}(p,t)$:
\begin{equation}
f(r,p,t)=f(r,p,0) - A(r,p) t;  \ \ r \ge R_{f}(p,t)
\label{eqn:linear_increase}
\end{equation}
i.e., the distribution function outside the flatness front falls linearly with time. More generally, we can solve for the flatness front $R_{\rm f}$ and the overall solution both inside and outside $R_{\rm f}$ via equations (\ref{eqn:analytic}) and (\ref{eqn:analytic_front}).

\begin{figure}
\includegraphics[width=8cm,trim=2cm 13cm 4cm 4cm]{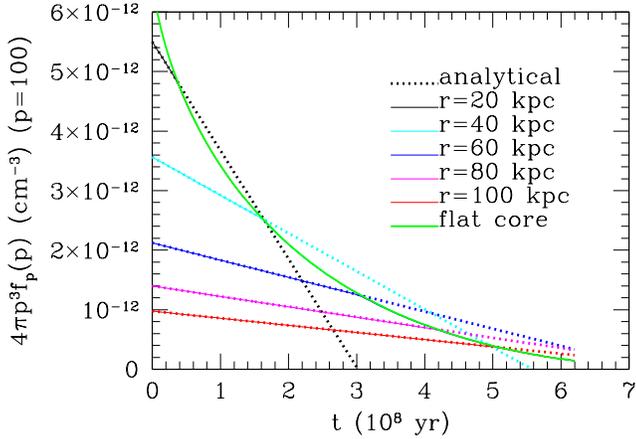}
\caption{CR densities at 100 GeV for Perseus if only the flux from CR streaming $\mathbf{F}_\tr{stream}$ (equation \eqref{eqn:F_stream}) is important. The dotted lines (bold curve) show the analytic solution for outside (inside) the flat front respectively; the solution initially follows the dotted curves until it intersects the green curve, when it follows the flat front solution. The solid lines show the simulation results, which match the analytic solution almost perfectly.}
\label{diffcompP1}
\end{figure}

\begin{figure}
\includegraphics[width=8cm,trim=2cm 10cm 4cm 4cm]{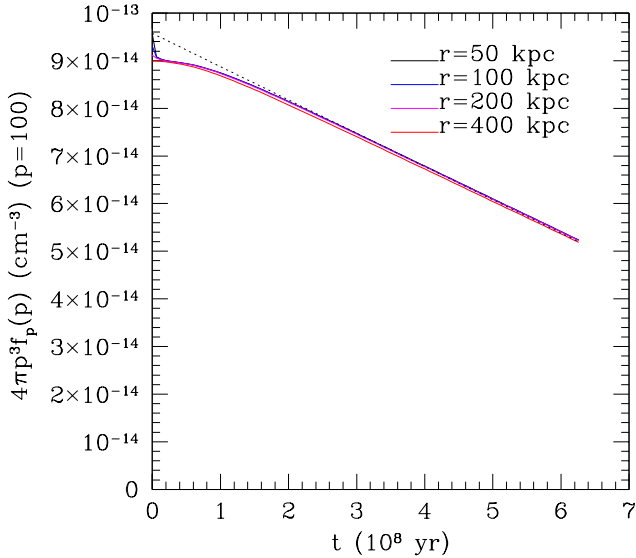}
\caption{The same as for Fig \ref{diffcompP1} but for Coma. Since the profile is already almost flat, this is a test of the $\dot{f}_\tr{p}=-3F_\tr{cr}(R_\tr{f})/R_\tr{f}=$ constant regime.}
\label{diffcompC1}
\end{figure}

To compare this analytic solution with our simulation we ran a simulation for Perseus and for Coma where only the diffusion term was used in \eqref{crevol}, and all other terms ignored. The resulting CR densities for Perseus can be seen in figure \ref{diffcompP1}. In this plot we show the CR density versus time at 100 GeV at a few select radii. The solid curves are the simulation, and the dotted lines are the analytic solution for a non-flat profile. The match is essentially perfect - the densities decrease at a constant rate (equation (\ref{eqn:linear_increase})) until the flatness front catches up to each radius. After this point the densities follow the same single curve corresponding to the evolution of the flat region. The same results for Coma show the agreement in the regime when the profile is already flat. In figure \ref{diffcompC1}, the CR profile in Coma is nearly flat to begin with. Before very long the profile is flat across the entire simulated space and the $\dot{f}_\tr{p}=-3F_\tr{cr}(R_\tr{f})/R_\tr{f}$ regime kicks in. Again, the agreement between simulation and analytic solution is very good. This implies that our regularization of the CR streaming term (which is needed to prevent unphysical oscillations with such flat profiles) is not so strong that it artificially changes the rate of diffusion.

\begin{figure}
\includegraphics[width=8cm,trim=2cm 10cm 4cm 4cm]{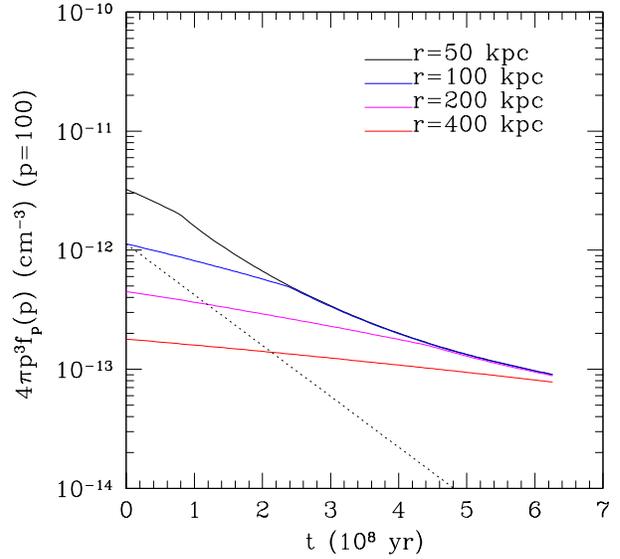}
\caption{CR densities for Perseus in the absence of any diffusion, i.e. only adiabatic losses are used. The dotted line represents the analytic solution \eqref{eqn:analytic_adiabatic} and should be compared to the blue line.}
\label{aa}
\end{figure}

{\bf Adiabatic expansion only.} We have:
\begin{equation}
{\mathbf F}_\tr{adia} = \frac{1}{3} p \frac{\partial f_\tr{p}}{\partial p} {\mathbf v}_\tr{A}
\label{eqn:flux_adia}
\end{equation}
Thus, unlike the preceding case, the flux depends on the distribution function $f_\tr{p}$ and hence is time-dependent. We can readily solve this in the approximation that $v_{\rm A} \propto \rho^{\alpha_{\rm B} - 0.5} \sim$ const (since it varies very weakly with radius), and $f_\tr{p} \propto p^{-\alpha}$, approximately independent of radius. Then, for $r \ge R_\tr{f}$, we have $\dot{f} = \nabla \cdot {\mathbf F}_\tr{adia} \approx -2\alpha v_\tr{A} f_\tr{p}/3r$, or:
\begin{equation}
f_\tr{p}(r,p,t) \approx f_\tr{p} (r,p,0) {\rm exp} \left( - \frac{2 \alpha v_\tr{A} t}{3 r} \right);  \ \ r \ge R_\tr{f}
\label{eqn:analytic_adiabatic}
 \end{equation}
Thus, outside the flatness front, the distribution function falls exponentially with time, with e-folding time of order the Alfv\'en crossing time (which becomes long at large radii). To solve for the evolution of the flatness front and the entire profile, we insert equation (\ref{eqn:analytic_adiabatic}) into equation (\ref{eqn:flux_adia}) and hence equation (\ref{eqn:analytic}) and (\ref{eqn:analytic_front}). Note that we are only required to evaluate the flux ${\mathbf F}_\tr{adia}$ for $r \ge R_{f}$, where equation (\ref{eqn:analytic_adiabatic}) is valid. We compare this analytic expression with a Perseus simulation that has no diffusion in Fig \ref{aa}. The dashed line depicts \eqref{eqn:analytic_adiabatic} for $r=100$ kpc and $p=100$. Although the fit isn't perfect, the simulated values do fall exponentially with time until the flatness front catches up, with e-folding time comparable to that determined from \eqref{eqn:analytic_adiabatic}. This is perhaps to be expected, since \eqref{eqn:analytic_adiabatic} assumes that the quantity $\alpha f_\tr{p}$ does not vary significantly with radius, which is not typically the case.
\label{lastpage}

These solutions allow us to understand the nature of the numerical solutions we previously obtained. Coma, where the initial profile is almost completely flat, is obviously in the $\dot{f}_\tr{p}=-3F(R_\tr{max})/R_{f}$ regime; moreover, ${\mathbf F}_{\rm stream}(R_{f}) \gg {\mathbf F}_{\rm adia}(R_{\rm f})$, since the latter scales with the (small) value of the distribution function at the outer boundary. The evolution of the flatness front in Perseus is more interesting. Initially, even though $v_\tr{D} - v_\tr{A} \sim {\cal O}(v_{A})$, adiabatic losses dominate, since the distribution function falls exponentially with time (rather than linearly with time, for streaming losses). However, ${\mathbf F}_\tr{adia} \propto f$ also falls exponentially with time, while ${\mathbf F}_\tr{stream}$ is independent of time. Thus, streaming losses will always dominate at late times. Equivalently, the velocity associated with adiabatic losses, $v_\tr{A}$, is constant with time, while the streaming velocity $v_{D} \propto 1/f_\tr{p}$ increases with time: as the number density of cosmic rays fall, the confining wave amplitude $\delta B/B$ falls, and cosmic rays can stream progressively faster.

\section{Conclusions}
\label{sect:conclusions}

Shocks generated during hierarchical structure formation are expected to accelerate cosmic rays via diffusive shock acceleration. These cosmic rays in turn interact hadronically with thermal nucleons to produce pions, which decay to produce relativistic electrons. Tracking these well-understood processes, and assuming magnetic fields given by Faraday rotation measurements, leads to predictions for radio halo emission consistent with those observed \citep{pfrommer08}. However, this model predicts that {\it every} cluster hosts a bright radio halo. This is at odds with the observed bimodality of cluster radio emission: the majority of clusters are radio-quiet, and an order of magnitude fainter than the radio-loud population \citep{feretti12,brown11}. Radio loudness is strongly associated with merger activity. For this reason, the turbulent re-acceleration model \citep{brunetti01,petrosian01}, where this association occurs naturally, is often favored. However, this still begs the question as to {\it why} hadronically induced radio emission is not omnipresent. All of the associated physics is well understood, and at face value the observations then require that CRp acceleration efficiencies be reduced by an order of magnitude below canonical values\footnote{A bimodality in cluster B-fields, with larger values during the turbulent, radio-loud state, appears inconsistent with cluster rotation-measure observations (\citet{bonafede11}, and references therein).}.

\citet{enslin11} took an important step forward when they suggested that CRp's could potentially stream super-Alfv\'enically, turning off radio halos. However, they assumed streaming speeds of order the sound speed $v_{D} \sim c_{\rm s}$ instead of calculating it\footnote{In fact, given their assumptions, we find that cosmic rays should only stream Alfv\'enically.}, and posited steady-state CR profiles that represent equilibria between outward streaming and inward turbulent advection, despite the long timescales for equilibration. In this paper, we attempt to place CR streaming in clusters on a more rigorous footing, by calculating the microphysical streaming speed as a function of plasma parameters in the self-confinement picture \citep{lerche67, kulsrud69, wentzel69, skilling71}. We then solve the time-dependent CR transport equation (albeit in 1D) to see how the radio luminosity evolves with time. Our conclusions are as follows:
\begin{itemize}
\item{CR streaming speeds depend on the source of wave damping. Non-linear Landau damping (e.g., \citep{felice01}, as assumed in \citet{enslin11}) is too weak to sufficiently inhibit wave growth, and $v_{\rm D} \sim v_{\rm A}$. However, if waves are instead damped by turbulent shear \citep{yan02,farmer04}, they can be sufficiently suppressed that super-Alfv\'enic streaming $v_{\rm D} \gg v_{\rm A}$ is possible. Moreover, $v_{\rm D}-v_{\rm A} \propto \gamma/n_{\rm CR}(>\gamma)$, (where $\gamma$ is the CR Lorentz factor) so that: i) higher energy cosmic rays stream more rapidly; ii) CR streaming speeds continually increase as $n_{\rm CR}$ declines due to streaming.}
\item{Streaming relative to the Alfv\'en wave frame can be incorporated into the CR transport equation via a diffusion term. For turbulent wave damping, the diffusion coefficient $\kappa \propto 1/\nabla f_\tr{p}$ (where $f_\tr{p}$ is the distribution function), so that remarkably $\nabla \cdot (\kappa \nabla f_\tr{p})$ is {\it independent} of $\nabla f_\tr{p}$. Thus, CRs can continue to stream unabated in giant radio halos (such as Coma) despite their fairly flat inferred CR profiles. Streaming is still sensitive to the {\it sign} of $\nabla f_\tr{p}$ (since CRs can only stream down a gradient), and for flat profiles we must implement numerical regularization \citep{sharma09} to ensure stable solutions. We test our solver for the CR distribution function against a code where CR mediated AGN heating is solved in the fluid approximation \citep{Guo08a}. The solutions are identical. Note that CR heating is unaffected by super-Alfv\'enic streaming, since it scales as ${\bf v}_{\rm A} \cdot \nabla P_{c}$ and $P_{c}$ is dominated by $\sim$GeV CRs, where streaming is Alfv\'enic.}
\item{CR transport is thus clearly modified by ICM turbulence. Besides its effects on wave damping, turbulence can also advect CRs so that they roughly trace the gas density profile, creating a centrally peaked CR distribution. For the mildly subsonic $v_{\rm s} \sim v_{\rm A} \sim 100 \, {\rm km \, s^{-1}}$ turbulence we assume, outward streaming dominates inward advection. Moreover, this trend {\it increases} with the amplitude of turbulence. It is therefore consistent with the flat inferred CR profiles in non cool-core clusters, which have generally stronger turbulent motions. Such a trend is hard to understand in scenarios where turbulence only draws CRs inward \citep{enslin11,zandanel12}.}
\item{We then perform numerical time-dependent calculations of CR streaming, assuming an initial profile consistent with radio observations at 1.4 GHz. We find that the radio luminosity falls by an order of magnitude in several hundred Myr, both in a prototypical radio mini-halo (Perseus) and giant radio halo (Coma). The latter effect is particularly interesting in light of the flat inferred CR profile, and arises {\it only} for turbulent damping of MHD waves; if only non-linear Landau damping is at play, the turn-off is slow. Indeed, the inferred flatness of the CR profile suggests that streaming has already been at play in these systems. We also build an analytic model which aids in physical understanding. Adiabatic losses dominate until the profile flattens, when diffusive losses dominate. The turn-off timescale in the later stage is set by the lowest value of the CR flux $F \propto B^{3/2}/L_{\rm MHD}$, generally at the cluster outskirts. The energy-dependence of CR streaming means that spectral curvature develops, and radio halos turn off more slowly at low frequencies, both consistent with observations \citep{brunetti08,brunetti12}. Streaming also rapidly diminishes the $\gamma$-ray luminosities at the $E_{\gamma} \sim 0.3-1$ TeV energies probed by imaging air Cerenkov telescopes (MAGIC, HESS, VERITAS), but not for the lower energies $E_{\gamma} \sim 0.1-3$ GeV probed by Fermi. The latter is therefore a more robust probe of the CR injection history.}

\end{itemize}

The primary contribution of this paper is a physical proof of principle for turning off hadronically induced emission. Our 1D streaming calculations by nature omit important details best clarified by 3D MHD simulations. Chief amongst these are the effects of magnetic topology. We have effectively assumed radial magnetic fields in our 1D calculations. Of course, magnetic topology greatly influences the true value of macroscopic transport coefficients. There is some evidence both from observations \citep{pfrommer10} and cosmological MHD simulations \citep{ruszkowski11a} that outside the core, magnetic fields are largely radial, driven either by cosmological infall, or the magneto-thermal instability (MTI; \citet{balbus00,parrish08}). Alternatively, turbulence could fully tangle magnetic fields \citep{ruszkowski10,ruszkowski11,parrish09}. CRs have to follow the same field lines that thermal particles do, albeit with a larger gyro radius\footnote{Interestingly, 100 GeV CRs have a mean free path due to wave-particle interactions $\lambda_{\rm CR} \sim (\delta B/B)^{-2} r_{\rm L} \sim 1-10 \, {\rm kpc} B_{\mu G}^{-1} \epsilon_{\rm 100 \, GeV} ([\delta B/B]/10^{-4})^{-2}$ which is similar to the electron collisional mean free path $\lambda_{e} \sim 6 \, {\rm kpc} \, T_{4 \, {\rm keV}}^{2} n_{i,-3}$.}. As long as cross-field diffusivity remains small, transport coefficients should scale similarly; in the limit of a fully tangled field with a coherence length significantly larger than the gyro radius, a random walk in 3D rather than 1D will reduce the diffusion coefficient $\kappa_{\rm p} \rightarrow \kappa_{p}/3$, just as it reduces the Spitzer-Braginskii value for thermal conductivity by a factor of 3. This will effectively increase all quoted timescales in this paper for pure streaming by a factor of $\sim 3$. In the future, it would be interesting to conduct fully self-consistent 3D MHD simulations which include CR streaming, motivated and guided by the estimates here. We have also incorporated the advective effects of gas motions only in the diffusive approximation. Coherent bulk motions due to mergers or sloshing could potentially have stronger effects. More light on the nature of ICM motions in radio-bright halos from Astro-H (e.g., \citet{zhuravleva12,shang12,shang12a}) will surely help. We are also agnostic as to the cause of radio halo turn-on, which is clearly related to gas motions stimulated by mergers, and could be due to turbulent reacceleration of seed CRe \citep{brunetti01,petrosian01}, inward advection of CRs from the cluster outskirts \citep{enslin11}, or perhaps have separate mechanisms for different classes of radio halos \citep{zandanel12}. Such issues await clarification from low frequency radio observations by LOFAR.

\section*{Acknowledgments}

We acknowledge NSF grant 0908480 and NASA grant NNX12AG73G for support. We are grateful to Christoph Pfrommer for sparking our interest in this topic and stimulating discussions. We also thank Gianfranco Brunetti, Torsten Ensslin, Anders Pinzke and Ellen Zweibel for helpful conversations. SPO also thanks the KITP (supported by NSF PHY05-51164), the Aspen Center for Physics (NSF Grant No. 1066293) and UCLA for hospitality, and the Getty Center for inspiring views, during the completion of this paper.

\bibliography{master_references}

%%%\appendix

\end{document}